\documentclass[preprint,sort&compress,12pt]{elsarticle}

\usepackage{amssymb}
\usepackage{amsthm}
\usepackage{amsmath}
\usepackage{mathtools}
\usepackage{mathrsfs}
\usepackage{algorithm}
\usepackage[algo2e]{algorithm2e}
\usepackage{algpseudocode}
\usepackage{array}
\usepackage{multirow}
\usepackage{listings}
\usepackage{tabu}
\usepackage{booktabs}
\usepackage{enumerate}
\usepackage{fullpage}
\usepackage{float}
\usepackage{xcolor}
\usepackage[colorinlistoftodos]{todonotes}
\usepackage{bbm}
\usepackage[colorlinks=true]{hyperref}
\usepackage{url}
\usepackage{textcomp}
\usepackage{gensymb}
\usepackage{soul}
\usepackage{graphicx}
\usepackage{subfigure}
\usepackage{float}
\usepackage{caption}
\usepackage{tikz}
\usetikzlibrary{shapes}

\theoremstyle{definition}

\theoremstyle{remark}

\biboptions{numbers,comma,round,square}
\graphicspath{ {./figs/} }

\journal{Elsevier}

\begin{document}
\begin{frontmatter}

 \title{Differentiable hybrid neural modeling for fluid-structure interaction}



\author[ndAME,ndECI,ndEnergy]{Xiantao Fan}
\author[ndAME,ndECI,ndEnergy,ndLucy]{Jian-xun Wang\corref{corxh}}

\address[ndAME]{Department of Aerospace and Mechanical Engineering, University of Notre Dame, Notre Dame, IN}
\address[ndECI]{Environmental Change Initiative, University of Notre Dame, Notre Dame, IN}
\address[ndEnergy]{Center for Sustainable Energy (ND Energy), University of Notre Dame, Notre Dame, IN}
\address[ndLucy]{Lucy Family Institute for Data \& Society, University of Notre Dame, Notre Dame, IN}
\cortext[corxh]{Corresponding author. Tel: +1 540 3156512}
\ead{jwang33@nd.edu}

\begin{abstract}
Solving complex fluid-structure interaction (FSI) problems, which are described by nonlinear partial differential equations, is crucial in various scientific and engineering applications. Traditional computational fluid dynamics based solvers are inadequate to handle the increasing demand for large-scale and long-period simulations. The ever-increasing availability of data and rapid advancement in deep learning (DL) have opened new avenues to tackle these challenges through data-enabled modeling. The seamless integration of DL and classic numerical techniques through the differentiable programming framework can significantly improve data-driven modeling performance. In this study, we propose a differentiable hybrid neural modeling framework for efficient simulation of FSI problems, where the numerically discretized FSI physics based on the immersed boundary method is seamlessly integrated with sequential neural networks using differentiable programming. All modules are programmed in JAX, where automatic differentiation enables gradient back-propagation over the entire model rollout trajectory, allowing the hybrid neural FSI model to be trained as a whole in an end-to-end, sequence-to-sequence manner. Through several FSI benchmark cases, we demonstrate the merit and capability of the proposed method in modeling FSI dynamics for both rigid and flexible bodies. The proposed model has also demonstrated its superiority over baseline purely data-driven neural models, weakly-coupled hybrid neural models, and purely numerical FSI solvers in terms of accuracy, robustness, and generalizability. 

\end{abstract}

\end{frontmatter}

\section{Introduction}
\label{sec:intro}
Fluid-structure interaction (FSI) is ubiquitous in various scientific and engineering applications, spanning from small-scale biological systems to large-scale industrial structures in civil or hydraulic engineering. Two-way coupled FSI is a complex, multiphysics problem due to inherent structural and fluid nonlinearities, as well as a wide range of spatiotemporal scales involved in the structure, fluid, and their interactions~\cite{boustani2021immersed, griffith2020immersed}. Computational modeling is important for predicting and controlling complex FSI dynamics. However, traditional computational models based on classic numerical techniques face great challenges in efficiently simulating FSI problems with sufficient accuracy and low computational cost. Although many computational fluid dynamics (CFD) techniques, such as the arbitrary Lagrangian-Eulerian~\cite{ramaswamy1987arbitrary, donea1982arbitrary}, level-set~\cite{sethian1999level}, and immersed boundary method~\cite{RN1234, peskin1972flow, fadlun2000combined}, have been developed to model interactions between fluid and solid dynamics and provide many physical insights, solving a series of FSI governing ordinary/partial differential equations (ODEs/PDEs) remains very expensive, making it infeasible to deal with real-time predictions or many-query applications, e.g., design optimization, inference, and uncertainty quantification. 

Rapid advancements in machine learning and ever-increased data availability have provided new opportunities to tackle these challenges. The use of deep neural networks (DNNs) for modeling complex nonlinear dynamics of fluid or FSI systems has been actively explored in recent years. By directly learning the input-output relationships in either high-dimensional or reduced-order space using data-driven techniques such as proper orthogonal decomposition (POD) and convolutional autoencoder methods, mapping functions have been identified for various scenarios, including flow past cylinders~\cite{fu2021data,guo2019data}, airfoil aerodynamics~\cite{bhatnagar2019prediction,zhu2019machine}, biological flows~\cite{du2022deep,zhang2022deep}, and RANS/LES closure problems~\cite{wang2017physics,pan2018data,maulik2019subgrid,bae2022scientific}, among others. To handle irregular domains with unstructured data, graph neural networks (GNNs) have been used to build fast neural simulators for fluid and FSI dynamics~\cite{pfaff2020learning,han2022predicting}. For example, a combination of GNNs and POD was used to learn the fluid and solid dynamics of a benchmark FSI case, vortex-induced vibration of an elastically-supported cylinder~\cite{gao2022quasi}. 

While data-driven models have shown promising results in modeling FSI systems, they have limitations, including the high cost of data acquisition and the potential lack of generalizability or robustness. Therefore, it is important to strike a balance between the use of data-driven models and physics prior to create hybrid models that take advantage of the strengths of both approaches. A hybrid deep learning model that combines data-driven techniques with prior knowledge of physics, such as governing equations and constraints, can be highly effective in improving sample efficiency and generalizability, particularly when dealing with small or sparse data. For FSI modeling, this hybridization can be achieved through two different ways: (1) the integration of DNNs with projection-based model reduction and constraints of fluid-structure coupling, and (2) the incorporation of governing equations into DNN's loss functions or neural architectures. This latter is also referred to as physics-informed machine learning~\cite{karniadakis2021physics}. 

Most existing data-driven models for FSI problems belong to the first category. Typically, the high-dimensional fluid and structure dynamics are reduced into low-dimensional latent spaces using techniques like POD or convolutional auto-encoders, and the latent fluid and structure dynamics are learned by separate DNNs~\cite{gupta2022hybrid}. Physical interface constraints, such as moving interfaces (solid-to-fluid coupling) and fluid forces (fluid-to-solid coupling), are often used to couple the fluid and structure DNNs, which can be represented using methods like level-set functions \cite{gupta2022hybrid, bukka2021assessment, von2021using}, immersed boundary method (IBM) masks~\cite{wandel2020learning}, or direct forcing terms~\cite{fang2022immersed}. By doing so, both the structural responses and the fluid dynamics can be learned in a consistent manner. In some cases, if only the structural responses are of interest, a more accurate numerical solver can be used to solve the solid dynamics, while the flow field or only fluid forces are learned by DNNs~\cite{minakowska2021finite}. Although this approach might seem similar to the hybrid neural models belonging to the second category that will be introduced shortly, the training of DNNs here is often conducted offline, decoupled from the numerical solver and physical constraints. These loosely-coupled hybrid methods often face significant stability issues~\cite{wu2019reynolds}, especially for predicting long-span spatiotemporal dynamics~\cite{um2020solver}.

The second category of hybrid data-driven approaches considered here is referred to as physics-informed machine learning (PIML). A key contribution to this category is physics-informed neural networks (PINNs), which incorporate physical prior, such as governing equations, conservation laws, and boundary conditions, as parts of the standard loss function to regularize neural network training~\cite{raissi2019physics}. PINNs have demonstrated their capabilities in various problems, including forward and inverse modeling of fluid flow \cite{sun2020surrogate,arzani2021uncovering}, solid mechanics~\cite{gao2022physics, lai2021structural}, and FSI problems~\cite{kharazmi2021inferring, tang2022transfer}. However, PINNs impose the physics constraints softly by reconstructing PDE residuals using either automatic differentiation (AD) or numerical discretization, making the loss landscape very difficult to optimize~\cite{krishnapriyan2021characterizing}. The success and failure modes heavily rely on the setting of hyperparameters that weigh each loss terms.

Recently, a new direction in PIML is to incorporate known physics into neural architecture~\cite{liu2022predicting}. This is inspired by the relationship between DNN architecture and differential equations~\cite{long2018pde,chen2018neural}. Liu et al.~\cite{liu2022predicting} proposed a PDE-preserved neural network (PPNN) that integrates discretized PDE operators from partially known governing physics into the DNN architecture through convolutional residual connection network (ConvResNet) blocks in a multi-resolution setting. Compared to purely black-kbox ConResNet and other state-of-the-art (SOTA) neural operators, PPNN has been demonstrated to have much better performance in terms of accuracy, speed, and generalizability for predicting various spatiotemporal physics. From the numerical modeling perspective, the PDE-preserving portion in PPNN can be interpreted as numerical differentiation and time stepping implemented as convolutional residual network blocks, which can be seamlessly integrated with any black-box neural network structures to form a hybrid neural solver. Thanks to differentiable programming~\cite{innes2019differentiable}, the gradient can be back-propagated through the entire neural solver, allowing efficient end-to-end, sequence-to-sequence training of the entire model as a whole. Lately, leveraging physics in their discretized forms based on classical numerical techniques, along the line of differentiable programming to construct a differentiable hybrid solver has been attracting increasing attention~\cite{um2020solver, kochkov2021machine, list2022learned, wright2022deep, akhare2023physics}. The differentiable programming framework has been recently applied in learning turbulence parametrization \cite{frezat2022posteriori, list2022learned,shankar2023differentiable}, where subgrid-scale closures or other correlations are learned based on \textit{a posteriori} criteria. End-to-end differentiable learning has enabled significant improvement of \emph{a posteriori} predictions compared to loosely-coupled hybrid DNN-closure that struggle with stability issues as mentioned above~\cite{wu2019reynolds}. The integration of DNNs to construct hybrid neural solvers poses significant challenges for most traditional legacy solvers due to the lack of AD capability and GPU compatibility. To overcome this issue, there are growing efforts to develop differentiable CFD solvers such as PhiFlow~\cite{holl2019learning}, Jax-CFD~\cite{kochkov2021machine}, and Jax-Fluids~\cite{bezgin2023jax}. Although the hybrid differentiable neural modeling paradigm has shown a great promise, this field is still in its early stages and requires further development, especially for modeling complex multi-physics problems such as two-way coupled FSI dynamics.

In this work, we present a differentiable hybrid neural model for predicting coupled fluid-structure dynamics, which integrate governing PDEs into deep learning architectures to accelerate FSI simulations while maintaining a balance between predictive accuracy and efficiency. The proposed hybrid neural model is constructed within the differentiable programming framework using JAX, a Python library designed for high-performance ML research and differentiable modeling. Specifically, a hybrid recurrent network unit of fluid-structure coupling is built based on a differentiable incompressible Navier-Stokes solver on very coarse grids combined with trainable convResNet blocks to ensure accurate fluid predictions. Meanwhile, the solid dynamics is directly solved on a high-resolution grid via superresolution operations, and coupled with fluid using IBM direct forcing to enforce two-way interference~\cite{yang2012simple}. These hybrid FSI network units are combined to form a sequential recurrent network based on Long Short-Term Memory (LSTM), allowing sequence-to-sequence (Seq2Seq) training and prediction. The outstanding performance of the proposed hybrid neural FSI model is demonstrated through extensive experiments on two benchmark cases, where we compare its accuracy, robustness, generalizability, and efficiency with other SOTA baseline methods. To the best of the authors' knowledge, this study represents the first attempt to integrate deep learning with numerical solvers for predictive FSI modeling within the differentiable programming paradigm. The paper is organized as follows: Section~\ref{sec:meth} outlines the overall methodology, including problem formulations and model designs. Section~\ref{sec:result} presents the results of our neural FSI model for two different benchmark cases, along with comparisons to basline methods. We discuss the robustness, generalizability, and speedup of the proposed hybrid neural model in Section~\ref{sec:discussion}, and conclude the paper in Section~\ref{sec:conclusion}.

\section{Methodology}
\label{sec:meth}
\subsection{Problem formulation} 
\label{sec:problem equation}
FSI problems can be described by a series of coupled PDEs that exhibit spatial-temporal nonlinearity. In this work, we concentrate on the interactions between incompressible fluid and solid dynamics to demonstrate the fundamental concepts of the proposed differentiable neural FSI simulation method. Specifically, the fluid dynamics is governed by the incompressible Navier-Stokes equations,
\begin{equation}
\label{eq:ns}
\begin{aligned}
\nabla\cdot\mathbf{u} = 0, \hspace{7em}&\mathbf{x}, t \in \Omega_f \times [0, T]\\
\frac{\partial{\mathbf{u}}}{\partial t} = -(\mathbf{u}\cdot\nabla){\mathbf{u}}+\nu\nabla^2\mathbf{u}-\frac{1}{\rho}\nabla{p}+\mathbf{f}, \hspace{3em} &\mathbf{x}, t \in \Omega_f \times [0, T]
\end{aligned} 
\end{equation}
where $t$ and $\mathbf{x}$ are time and Eulerian space coordinates, respectively; fluid velocity $\mathbf{u}(t, \mathbf{x})$ and pressure $p(t, \mathbf{x})$ are both spatiotemporal functions defined in $\Omega_f \subset \mathbb{R}^2$; $\rho$ and $\nu$ represent fluid density and kinematic viscosity of the fluid, respectively; The solutions for velocity and pressure are uniquely determined by given initial and boundary conditions (IC/BCs).

The interaction between fluid dynamics and deformable structures is described by immersed boundary method (IBM), wherein the effect of solid dynamics is imposed on fluid using direct forcing term $\mathbf{f}$~\cite{RN1234}, satisfying the non-slip wall conditions along the interface $\Gamma \in \Omega_f \cap \Omega_s$. The structure dynamics can be described by the following equation in Lagrangian coordinates,
\begin{equation}
\label{eq: equation of solid}
\begin{aligned}
\mu_s \frac{\partial^2 \mathbf{w}}{\partial t^2}+EI \frac{\partial^4 \mathbf{w}}{\partial \mathbf{X}^4}=\mathbf{q}, \hspace{3em}&\mathbf{w}, t \in \Omega_s \times [0, T]
\end{aligned} 
\end{equation} 
where $\mathbf{X}$ is Lagrangian space coordinates; $\mathbf{w}(t,\mathbf{X})$ represents the dynamic response of structure, defined in $\Omega_s \subset \mathbb{R}^2$; $\mu_s$ is mass per unit length; $EI$ is flexibility; $\mathbf{q}(t,\mathbf{X})$ represents forces induced by the flow. In general, structures can generally be classified into two types, rigid bodies and flexible bodies. For two dimensional problems considered here, a flexible body can be viewed as an assembly of rigid beam elements based on Euler-Bernoulli beam theory~\cite{RN1208}. The transformation between the Eulerian (fluid) and Lagrangian (structure) variables can be realized by the Dirac delta function~\cite{peskin1972flow, uhlmann2005immersed},
\begin{equation}
\label{eq:interpolation}
\begin{aligned}
    \mathbf{\phi}(\mathbf{x},t)=\int_{\gamma_\mathbf{X}} \mathbf{\Phi}(\mathbf{X},t)\delta(\mathbf{x}-\mathbf{X})d\gamma_\mathbf{X},  \hspace{3em} &\gamma_\mathbf{X} \in \Gamma \\
    \mathbf{\Phi}(\mathbf{X},t)=\int_{\gamma_\mathbf{x}} \mathbf{\phi}(\mathbf{x},t)\delta(\mathbf{x}-\mathbf{X})d\gamma_\mathbf{x},  \hspace{3em} &\gamma_\mathbf{x} \in \Gamma 
\end{aligned}
\end{equation}
where $\mathbf{\Phi}$ and $\mathbf{\phi}$ represent variables in Lagrangian and Eulerian coordinates, respectively; $\delta$ denotes Dirac delta interpolation function; $\gamma_\mathbf{x}$ and $\gamma_\mathbf{X}$ correspond to the set of Eulerian grids and Lagrangian grids in the immersed interface boundary $\Gamma$, respectively.

Traditionally, finite difference (FDM) or finite volume method (FVM) are commonly used to solve Eq.\ref{eq:ns}, while finite element method (FEM) is used to solve Eq.\ref{eq: equation of solid}. However, the computational cost of traditional FSI solvers can be very high, promoting the development of a differentiable hybrid neural solver. By integrating discretized FSI physics (Eqs.\ref{eq:ns}-\ref{eq: equation of solid}) into deep neural architectures, this approach aims to achieve efficient and reliable FSI simulations.   


\subsection{Differentiable hybrid neural FSI modeling}
\begin{figure}[!t]
\centering
\includegraphics[width=\textwidth]{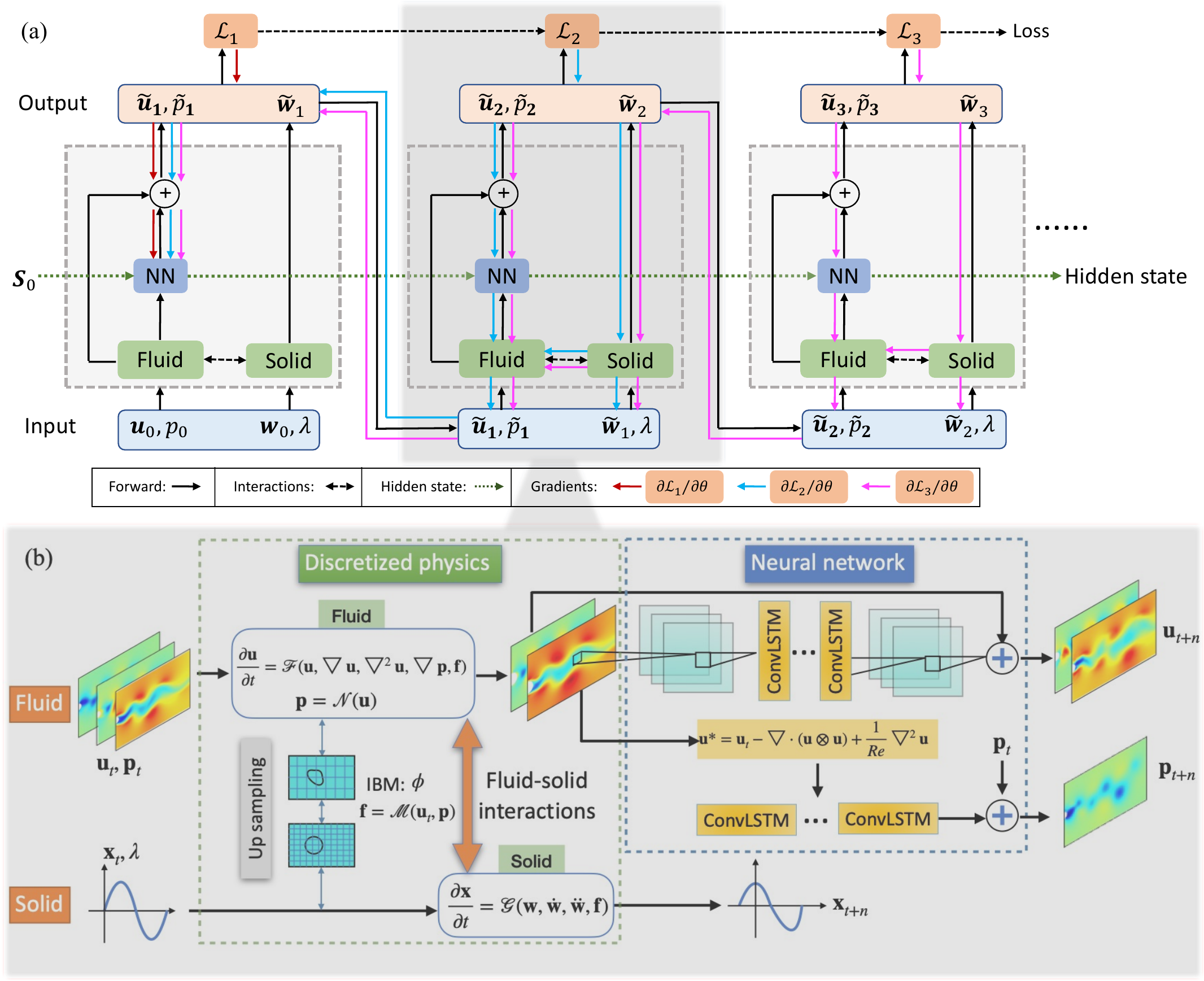}
\caption{(a) Overview of the differentiable hybrid neural architecture for FSI simulation, where `Fluid' and `solid' denote the governing PDEs of fluid and solid dynamics, respectively; `NN' represents trainable neural structures; The overall hybrid neural model is formulated as a LSTM sequence network based on FSI physics-integrated recurrent units. (b) The detailed schematic of the FSI-physics-integrated recurrent unit.}
\label{fig:overview scheme}
\end{figure}

The proposed differentiable hybrid neural FSI model is built upon differentiable programming, extending traditional deep learning frameworks to incorporate physical principles more effectively. The overview of the proposed hybrid neural architecture is depicted in Fig.~\ref{fig:overview scheme}(a), 
which is designed as a LSTM sequence net consisting of a series of recurrent units that integrate FSI physics. As detailed in Fig.~\ref{fig:overview scheme}(b), the physics-integrated recurrent unit encode the FSI physics, i.e., governing PDEs in discretized form, as non-trainable part of the neural architecture, seamlessly combined with trainable portion to form the entire LSTM sequence net. All modules are programmed in JAX, where the automatic differentiation (AD) enables the gradient back-propagation over the entire program, allowing the entire hybrid model to be trained in a Seq2Seq manner similar to classic sequence neural networks.     

Specifically, for each recurrent unit of the current step $t$, the input are the predicted velocity ($\mathbf{\mathbf{u}_{t}}$), pressure ($p_{t}$), and structure response ($\mathbf{w}_t$) from the previous learning step. The input initially pass through the physics-encoded portion with non-trainable spatial convolutions determined by the PDE operators from the governing FSI equations, which can be viewed as classic numerical time stepping implemented as convolution layers on low-resolution grids. The information then proceeds through additional trainable convolution layers, formulated as ConvLSTM with residual connections. In the physics-encoded portion, the computationally-expensive Poisson equation is entirely replaced by trainable convolutions to avoid costly self-iterations, and the velocity is corrected by the predicted pressure, following the fractional step procedure~\cite{chorin1968numerical}. It is worth mentioning that learning step ($\delta t$) can be much greater than numerical time step determined by Courant–Friedrichs–Lewy (CFL) condition, enabling fast simulation speeds. The governing equations for solid motions are directly solved due to their low degrees of freedom. Fluid and solid interactions occur through super-resolution operations, indicating that the mask function $\phi$ and direct forcing $\mathbf{f}$ of the IBM are executed in an super-resolved high-resolution grid (see Fig.\ref{fig:overview scheme}(b)). 

\subsubsection{Encoding fluid physics in discretized form}
\label{sec:encode-fluid}
The fluid physics, i.e., Navier-Stokes equations, is encoded in its FVM-based discretized form using fractional step method. Specifically, with a forward Euler time advancement, the intermediate velocity $\mathbf{u}^*$ is estimated based on the discretized momentum equations, 
\begin{equation}
\mathbf{u}^{*} = -\mathbf{u}^t -{\Delta t}\left[\nabla\cdot(\mathbf{u}^{t}\otimes\mathbf{u}^{t})+\nu \nabla^2\mathit{\mathbf{u}^{t}}+\mathbf{f}^{t}\right],
\end{equation}
where $\Delta t$ is the time step, $\otimes$ denotes the tensor product operation, and the direct IBM forcing term can be computed as follows,
\begin{equation}
\mathbf{f}^{t}=\varepsilon^t(\mathbf{x}) \left[\nabla\cdot(\mathit{\mathbf{u}^t}\otimes\mathit{\mathbf{u}^t})-\nu\nabla^2\mathit{\mathbf{u}^t}+\frac{\mathbf{u}_s^{t}-\mathbf{u}^{t}}{\Delta t}\right],
\end{equation}
where $\mathit{\mathbf{u}_s^t}$ is the target velocity of the immersed interaction boundary, which is interpolated by the Dirac delta function from structure velocity (Eq. \ref{eq:interpolation}); $\varepsilon^t(\cdot)$ represents the volume-of-solid function, with $\varepsilon^t(\mathbf{x}_s)$ = 1 for solid cells $\mathbf{x}_s \in \Omega_s$ and $\varepsilon^t(\mathbf{x}_f)$ = 0 for fluid cells $\mathbf{x}_f \in \Omega_f$. After that, the pressure term should be obtained by solving the Possion equation,
\begin{equation}
\nabla^2p^{t+1}=\frac{\rho \nabla \cdot [\mathbf{u}^{*}- \mathbf{u}^{t+1}]}{\Delta t}
\label{eq: initial possion}  
\end{equation} 
Incompressible condition $\nabla \cdot \mathbf{u}^{t+1} = 0$ is valid in the fluid domain $\Omega_f$ but does not hold in the solid domain $\Omega_s$ as the flow solution inside $\Omega_s$ is not physical. Therefore, the incompressible condition should be corrected for the solid domain,
\begin{equation}
\label{eq:correction of incompressible}
\nabla \cdot \mathbf{u}^{t+1}= \nabla \cdot \left[ \varepsilon^t(\mathbf{x})(\mathbf{u}_s^{t+1}-\mathbf{u}_c^{t+1}) \right] \approx \nabla \cdot \left[ \varepsilon^t(\mathbf{x})(\mathbf{u}^{*}-\mathbf{u}_c^{t}) \right] 
\end{equation}
where $\mathit{\mathbf{u}_c^t}$ is the exact velocity of the immersed solid. Consequently, the modified Poisson equation for the fluid-structure interface region is given as,
\begin{equation}
\nabla^2p^{t+1} = \frac{\rho\nabla \cdot \left[(1-\varepsilon^t)\mathbf{u}^{*}+\varepsilon^t \mathbf{u}_c^{t}\right]}{\Delta t}, \label{eq: pressure projection}   
\end{equation}
Note that when $\varepsilon=0$, the above equation reduces to the standard Possion pressure equation for pure fluid. However, instead of solving the Poisson equation, which is often computational intensive, trainable neural networks are employed to accelerate the process. In particular, trainable ConvLSTM layers are used to learn the Poisson operator, adhering to the same physical formulation as Eq.~\ref{eq: pressure projection},
 \begin{equation}
p^{t+1}=\mathscr{F}_{conv_1}[(1-\varepsilon^t)\mathbf{u}^{*}+\varepsilon^t \mathbf{u}_c^{t}, \mathbf{s}_{conv_1}^t, \boldsymbol{\lambda}; \boldsymbol{\theta}_1] + p^t, \label{eq:conv for pressure}
\end{equation}
where $\mathscr{F}_{conv_1}[\cdot]$ represents ConvLSTM neural function with hidden state $\mathbf{s}_{conv_1}^t$ at time $t$ and trainable parameters $\boldsymbol{\theta}_1$; $\boldsymbol{\lambda}$ represent physical parameters, e.g., material properties of the structure. Hereinafter, the output hidden state $\mathbf{s}_{conv_1}^{t+1}$ is omitted for simplicity.
The intermediate next step fluid velocity output from the physics-encoded part is obtained as follows,
\begin{equation}
\mathbf{\widetilde{u}}^{t+1}=\mathbf{u}^{*}  -\frac{\Delta t}{\rho}\nabla{p}^{t+1} \label{eq: final low-res velocity}
\end{equation}
The intermediate output $\mathbf{\widetilde{u}}^{t+1}$ pass through another trainable ConvLSTM block with residual connections to obtain the final velocity output $\mathbf{u}^{t+1}$, 
\begin{equation}
\mathbf{u}^{t+1}=\mathscr{F}_{conv_2}(\mathbf{\widetilde{u}}^{t+1},\mathbf{u}^{t},\varepsilon^{t+1}, \varepsilon^{t}, \mathbf{s}_{conv_2}^t, \boldsymbol{\lambda}; \boldsymbol{\theta}_2)+\mathbf{\widetilde{u}}^{t+1}, \label{eq: final velocity}
\end{equation}
where the volume-of-solid function $\varepsilon$ is used as a mask to provide the location information of the solid boundary; the ConvLSTM neural function $\mathscr{F}_{conv_2}$ parameterized with trainable parameters $\boldsymbol{\theta}_2$ takes the fluid and solid information from the PDE-encoded portion, as well as the hidden state $\mathbf{s}_{conv_2}^t$ from previous steps, to predict the next-step velocity. 

\subsubsection{Encoding solid dynamics in disretized form}
\label{sec:encode-solid}
The solid dynamics is directly encoded in the hybrid neural model using classic numerical techniques. Using the standard Galerkin method~\cite{zienkiewicz2005finite}, the solid deformation equation (Eq.~\ref{eq: equation of solid}) can be discretized as ordinary differential equations,
\begin{equation}
\mathbf{M} \frac{\partial^{2}\mathit{\mathbf{w}}}{\partial t^{2}}+ \mathbf{C} \frac{\partial \mathit{\mathbf{w}}}{\partial t}+\mathbf{Kw}=\mathbf{Q} \label{eq: vibration equation}
\end{equation}
where $\mathbf{w}=[\mathbf{w}_1, \mathbf{w}_2,\cdots,\mathbf{w}_n]^T$ represents the dynamic response vector for $n$ structural nodes, with each node's response comprised of its x- and y- displacement components $w_x$ and $w_y$, i.e., $\mathbf{w}_i=[w_{ix}, w_{iy}]^T$. The global mass, damping, and stiffness matrices are denoted by $\mathbf{M}, \mathbf{C},$ and $\mathbf{K}$, respectively. Lastly, $\mathbf{Q}=[\mathbf{q}_1, \mathbf{q}_2,\cdots,\mathbf{q}_n]^T$ represents the vector of fluid forces.  
The fluid forces $\mathbf{q}_i = [q^D_i, q^L_i]^T$ at $i^\mathrm{th}$ Lagrangian node can be computed by integrating the direct forcing $\mathbf{f}$ along the neighboring Eulerian nodes, 
\begin{equation}
q^D_{i}=-\int_{\gamma_\mathbf{x}}{f}_{x}\delta(\mathbf{x}-\mathbf{X})d\gamma_\mathbf{x} \cong-\sum_{\mathbf{x}_{i} \in \gamma_\mathbf{x}} \hat{{f}}_{x}(\mathbf{x}_{i})\Delta x \Delta y \label{eq:Fd}
\end{equation}
\begin{equation}
q^L_{i}=-\int_{\gamma_\mathbf{x}}{f}_{y}\delta(\mathbf{x}-\mathbf{X}) d \gamma_\mathbf{x}\cong-\sum_{\mathbf{x}_{i} \in \gamma_\mathbf{x}}\hat{{f}_{y}}(\mathbf{x}_{i})\Delta x \Delta y
\label{eq:Fl}
\end{equation}
where $q^D_{i}$ and $q^L_{i}$ represents the drag force and lift on the $i^\mathrm{th}$ node in the $x$ and $y$ direction, respectively. The terms ${f}_x=\mathbf{f} \cdot \mathbf{n}_x$ and ${f}_y=\mathbf{f} \cdot \mathbf{n}_y$ represent the $x$ and $y$ components of the Eulerian forcing term, with unit directional vector $\mathbf{n}_x$ and $\mathbf{n}_x$ along the $x$ and $y$ axes, respectively; $\hat{\Box}$ indicates the spreading of force from the Eulerian to the Lagrangian grid based on Eq.\ref{eq:interpolation}.
The time schemes used for solving Eq.\ref{eq: vibration equation} are the fourth-order Runge-Kutta method for rigid bodies~\cite{FAN2020104136} and the Newmark-beta method for flexible bodies~\cite{kamakoti2002computational}, respectively. In the current work, weak coupling between the fluid and structure is employed. The detailed forms of Eq.\ref{eq: vibration equation}-\ref{eq:Fl} for rigid and flexible bodies can be found in \ref{sec:append-equation}.

\subsection{Differentiable programming enabled Seq2Seq training}
The entire differentiable hybrid neural model can be trained in a Seq2Seq manner as a whole similar to classic sequence networks, e.g., recurrent neural networks or transformer, since all the components are constructed using differentiable programming. For Seq2Seq learning, the loss function accounting for both fluid and solid dynamics is defined as follows,   
\begin{equation}
\mathcal{L}(\boldsymbol{\theta}_1, \boldsymbol{\theta}_2) = \alpha \left[ \mathcal{L}_p (\boldsymbol{\theta_1}) + \mathcal{L}_v (\boldsymbol{\theta_2})\right] + \beta \mathcal{L}_s(\boldsymbol{\theta_1}, \boldsymbol{\theta_2})
\label{eq:loss}
\end{equation}
where $\boldsymbol{\theta_1}$ and $\boldsymbol{\theta}_2$ are trainable parameters of ConvLSTM layers; $\mathcal{L}_p$, $\mathcal{L}_v$, and $\mathcal{L}_s$ represent the loss components associated with fluid pressure, velocity, and solid response, respectively. The detailed expression of each loss component is given as follows, 
\begin{equation}
\mathcal{L}_p(\boldsymbol{\theta}_1) = \frac{1}{N} \sum_{t=0}^{N-1} \lVert p^{t}+\mathscr{F}_{conv_1}[(1-\varepsilon^t)\mathbf{u}^{*}+\varepsilon^t \mathbf{u}_c^{t}, \mathbf{s}_{conv_1}^t, \boldsymbol{\lambda}; \boldsymbol{\theta}_1] -\mathbf{p}^{t+1}_{d} \rVert_{L_2}^{2} 
\label{eq:pressure_loss}
\end{equation}

\begin{equation}
\mathcal{L}_v(\boldsymbol{\theta}_2) = \frac{1}{N} \sum_{t=0}^{N-1} \lVert \widetilde{\mathbf{u}}^{t+1}+\mathscr{F}_{conv_2}(\mathbf{\widetilde{u}}^{t+1},\mathbf{u}^{t},\varepsilon^{t+1}, \varepsilon^{t}, \mathbf{s}_{conv_2}^t, \boldsymbol{\lambda}; \boldsymbol{\theta}_2) -\mathbf{u}^{t+1}_{d} \rVert_{L_2}^{2} 
\label{eq:vel_loss}
\end{equation}

\begin{equation}
\mathcal{L}_s(\boldsymbol{\theta}_1, \boldsymbol{\theta}_2) = \frac{1}{N} \sum_{t=0}^{N-1}\lVert \mathbf{w}^{t+1} - \mathbf{w}^{t+1}_{d} \rVert_{L_2}^{2} 
\label{eq:structure_loss}
\end{equation}
where $\lVert \cdot \rVert_{L_2}$ represents the L2 norm, $N$ denotes the total number of rollout time steps, and the subscript ${\Box}_d$ indicates the labeled data. $\alpha$ and $\beta$ are the weighting parameters of the fluid and solid loss portions, respectively. The weights of these two losses chosen to be comparable in magnitude with each other to balance the contributions of the fluid and solid dynamics. The entire differentiable neural model is trained by optimizing the total loss over a long rollout trajectory using stochastic gradient descent, 
\begin{equation}
\boldsymbol{\theta}_1^{*}, \boldsymbol{\theta}_2^{*} =\operatorname*{argmin}_{\boldsymbol{\theta}_1,\boldsymbol{\theta}_2} \bigg[ \alpha \left( \mathcal{L}_p (\boldsymbol{\theta_1}) + \mathcal{L}_v (\boldsymbol{\theta_2})\right) + \beta \mathcal{L}_s(\boldsymbol{\theta_1}, \boldsymbol{\theta_2}) \bigg]\label{eq:minumloss}     
\end{equation}

The neural model unrolling over multiple time steps during training can significantly enhance inference performance over long trajectories, in terms of both accuracy and stability~\cite{um2020solver, kochkov2021machine}. The multi-step rollout in the training process is enabled by differentiable programming, which allows the gradient to be backpropagated over the entire rollout trajectory. However, naively applying Seq2Seq training of a hybrid neural model via differentiable programming may introduce some challenges: (1) the hybrid neural solver is susceptible to instability, particularly at the beginning of the training phase, due to nonphysical initialization of trainable parameters, which makes the discretized physics numerically unstable; (2) training over a long trajectory at once demands substantial GPU memory, as the gradients of the nested functions, resulting from a large number of model rollout steps, need to be stored. To address these issues, the entire training trajectory is divided into several subchains, which are connected head-to-tail. As demonstrated in \ref{sec:trajectory}, the gradients of each subchain can be combined to obtain the final gradients for updating trainable parameters using the chain rule. The hidden states of ConvLSTM are preserved and transferred between subchains, ensuring continuity and maintaining model coherence during training. The two neural networks introduced here for velocity correction and pressure prediction have the same architectures and hyperparameters, which can be found in \ref{sec: Hyperparameters} for details.

\section{Numerical results}
\label{sec:result}

\subsection{Case settings}
\label{sec:case}
In this study, we demonstrate the performance of the proposed differentiable hybrid neural FSI solver using two representative FSI benchmark cases: vortex-induced vibration (VIV) of a rigid body and flow-induced deformation (FID) of a flexible body. Firstly, the model equations described in Sec.~\ref{sec:problem equation} are directly solved by classic numerical FSI solver on a sufficiently high spatio-temporal resolution without incorporating any deep learning modules. This process aims to generate high-fidelity data for both training and validation of the proposed neural FSI model.
\begin{figure}[!ht]
\centering
\includegraphics[width=8cm]{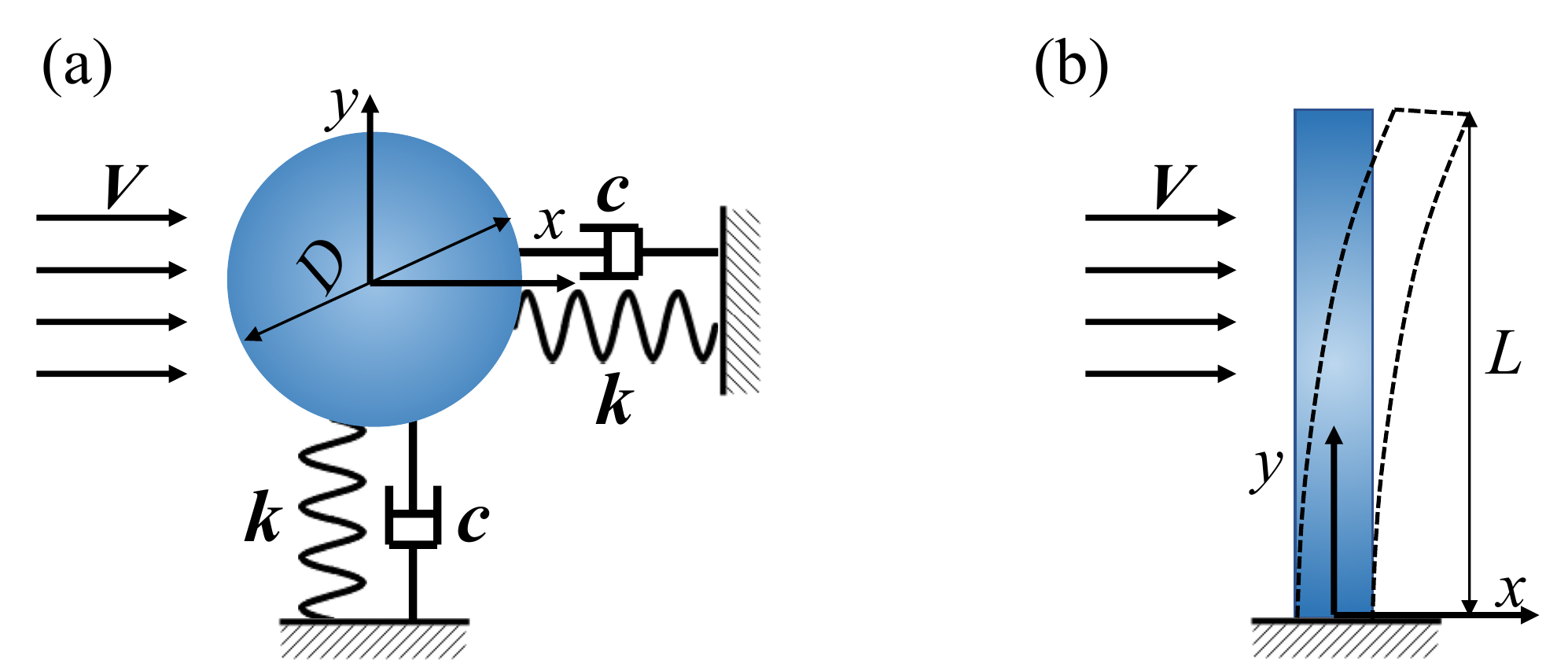}
\caption{(a) Vortex-induced vibration of a rigid cylinder; (b) Flow-induced deformation of a flexible plate.}
\label{fig:case settings}
\end{figure}
For the VIV of a rigid body, as shown in Fig.~\ref{fig:case settings}(a), the problem can be simplified as a typical spring-damping ($k-c$) system, as described in~\cite{FAN2020104136}. On the other hand, the flexible body in Fig.~\ref{fig:case settings}(b) can be modeled as a beam, with its deformation solved using the Euler-Bernoulli beam theory~\cite{Bauchau1208}. The parameters used in both cases for data generation are summarized in Tab.~\ref{tab:simulation parameter}, where most parameters are non-dimensional. 
\begin{table}[!t]
    \centering
    \footnotesize
    \caption{Summarize of case parameters for data generation.}
    \begin{tabular}{c c c c c c}
    \toprule[1.5pt]        
         Parameter & mass ratio ($m^*$) & natural frequency ($f_n$) & damping ($c$) & - & Re \\ \midrule
         Rigid body & 2 & 0.1-0.3 & 0 & - & 150 \\ \midrule \midrule
         Parameter & mass ratio($m^*$) & Young's modulus ($E$/Pa) & damping ratio & Poisson's ratio & Re \\ \midrule 
         Flexible body& 2.75 & $1\times 10^8$ & 0 & 0.3 & 150 \\ 
    \bottomrule[1.5pt]  
    \end{tabular}
    \label{tab:simulation parameter}
\end{table}
For example, the non-dimensional system mass is defined as mass ratio $m^*$,
\begin{equation}
m^*=\frac{4m}{\pi \rho_s B^2}, \label{eq:mass ratio}     
\end{equation}
where $m$ represents the mass, $B$ is the characteristic length with $B = D$ for the cylinder and $B = L$ for the flexible beam, $\rho_s$ denotes material density. The natural frequency $f_n$ is defined as,
\begin{equation}
f_n=\sqrt{\frac{k}{m}}, \label{eq:frequency}     
\end{equation}
where $k$ is spring stiffness. The Reynolds number is defined as $Re = VB/\nu$, where $V$ is the uniform free-stream velocity magnitude. To facilitate comparison with previous simulation and experimental results, the free-stream velocity is often normalized by characteristic length and natural frequency, which is known as reduced velocity $U_r$,
\begin{equation}
U_r=\frac{V}{Bf_n} \label{eq:reduced velocity}     
\end{equation}

\begin{figure}[!t]
\centering
\includegraphics[width=0.9\textwidth]{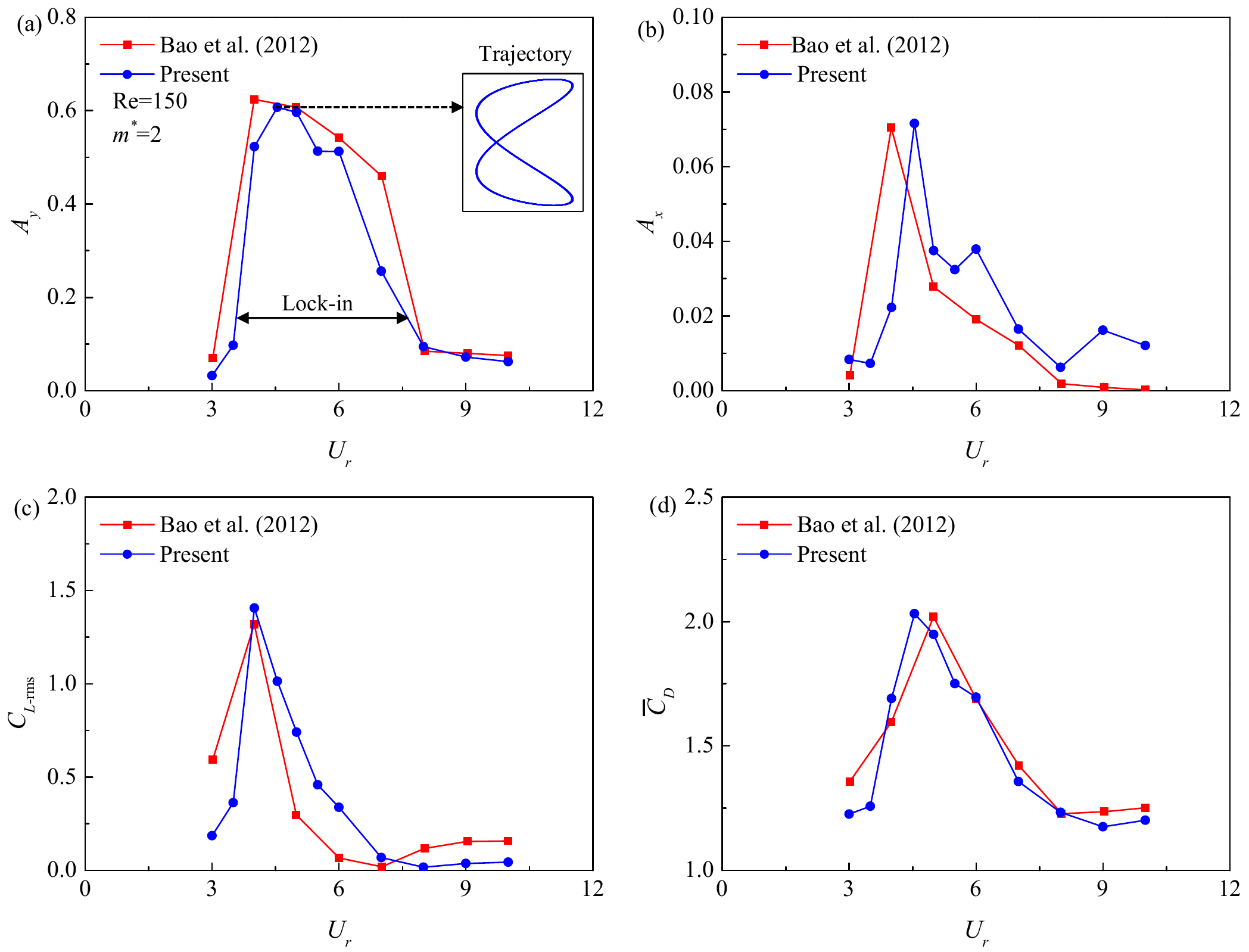}
\caption{Numerical results of the vortex-induced vibration (VIV) responses of a rigid cylinder: vibration amplitudes ($A_x, A_y$) in (a) transverse direction and (b) streamwise direction; (c) root-mean-square lift force $C_{L_rms}$; (d) average drag force $\bar{C_D}$. The reference data is from \cite{bao2012two}. The typical ``8"-shaped trajectory and ``lock-in" phenomenon are accurately simulated.}
\label{fig:2-VIV responses}
\end{figure}

\begin{figure}[!t]
\centering
\includegraphics[width=0.9\textwidth]{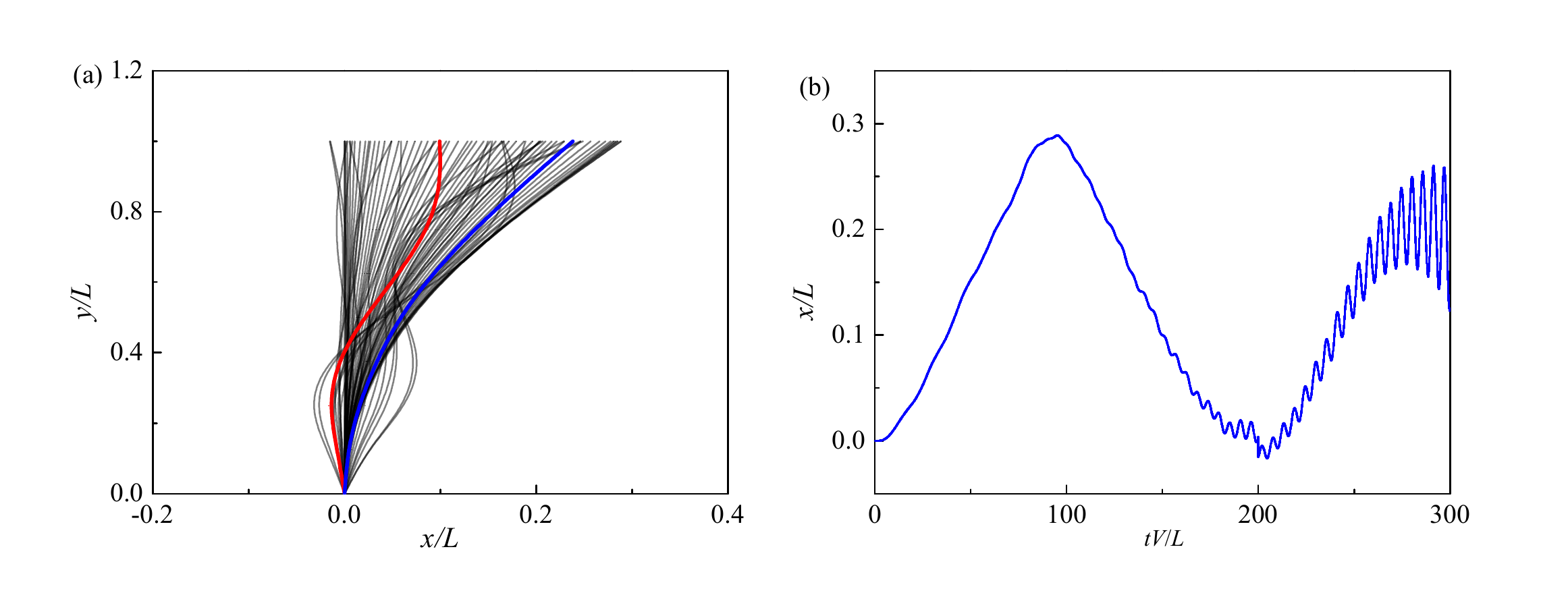}
\caption{Numerical results of the flow-induced deformation (FID) responses of a vertical flexible plate: (a) vibration shapes of the plate, shifting from the first mode (blue line) to the third mode (red line); (b) temporal deformations of the free-end of plate.}
\label{fig:FID of flexible responses}
\end{figure}

The numerical results of the VIV of a rigid body and the PID of a flexible body are demonstrated in Figs.~\ref{fig:2-VIV responses} and~\ref{fig:FID of flexible responses}, respectively. 
As shown in Fig.~\ref{fig:2-VIV responses}, our simulated results of the VIV response over various reduced velocities agree well with the previous numerical results obtained by Bao et al.~\cite{bao2012two}. The typical ``8"-shaped trajectory and ``lock-in" phenomenon are accurately captured in our present simulation.
Fig.\ref{fig:FID of flexible responses} displays the simulated results of the deformations of a vertical flexible plate. The oscillation shape in Fig.\ref{fig:FID of flexible responses}(a) transitions from the first mode to the third mode, indicating that the elasticity of the flexible structure has been captured in our simulation. From the temporal deformations of the free end of the plate shown in Fig.\ref{fig:FID of flexible responses}(a), we can observe that the vertical cantilever plate within the uniform flow initially experiences stable deformation in the streamwise direction, and then begins to oscillate due to vortex shedding in the wake, as reported in~\cite{tian2014fluid, connell2007flapping}. The numerical results obtained from the two distinct cases demonstrate that our differentiable numerical solver is capable of accurately simulating FSI physics with a sufficiently high spatio-temporal resolution. The high-resolution numerical simulations will be utilized for data generation and validation studies, supporting training and evaluation of the proposed hybrid neural model in a wide range of FSI scenarios.      

\subsection{Baseline models for comparison}
To demonstrate the advantages of our proposed differentiable hybrid neural model for FSI simulations, we compare it with two baseline learning models: a purely data-driven neural network model and a weakly-coupled hybrid neural model, as illustrated in Fig.~\ref{sfig:base line}. Additionally, we include a comparison with a purely numerical FSI solver that uses the same spatio-temporal resolution as the learning models.

The purely data-driven model, shown in Fig.~\ref{sfig:base line}(a), is a black-box sequence-to-sequence neural network that does not incorporate any physics prior knowledge. The fluid and solid dynamics are modeled by two separate recurrent neural networks: $NN_1$, a ConvLSTM with the same trainable neural architecture and parameters as the proposed hybrid neural model, and $NN_2$, a multi-layer perceptron (MLP)-based LSTM network. Notably, the fluid network $NN_1$ and solid network $NN_2$ are coupled based on IBM, where the IBM mask information is provided by the solid net $NN_2$ to the fluid net $NN_1$ to enforce fluid-structure interactions.

The weakly coupled hybrid neural model has the same neural architecture and trainable parameters as the proposed differentiable model. As depicted in Fig.~\ref{sfig:base line}(b), it also retains the physics of fluid and solid dynamics, discretized in the same way as described in Sec.~\ref{sec:encode-fluid} and Sec.~\ref{sec:encode-solid}. However, the key difference between this baseline with the proposed model is that the physics-encoded portions are not differentiable. Consequently, the gradients are halted in each trainable network component and cannot be back-propagated through the entire model rollout trajectory, making the model equivalent to an offline hybrid net trained in a teacher-forcing manner. Namely, this model is also informed by the governing PDEs of FSI dynamics, but in a weakly coupled manner. This baseline model falls into the first category of hybrid models as discussed in Sec.\ref{sec:intro}, representing the hybridization of a numerical solver with an offline trained neural network correction from solver point of view.

Henceforth, the two baseline learning models are referred to as ``purely data-driven" and ``weakly coupled," respectively. The model that only preserves the discretized physics portion without any trainable components is equivalent to a traditional explicit IBM-based FSI solver and is subsequently denoted as the ``pure solver." 


\subsection{Vortex-induced vibration (VIV) of a rigid body}
\begin{figure}[!t]
\centering
\includegraphics[width=1.0\textwidth]{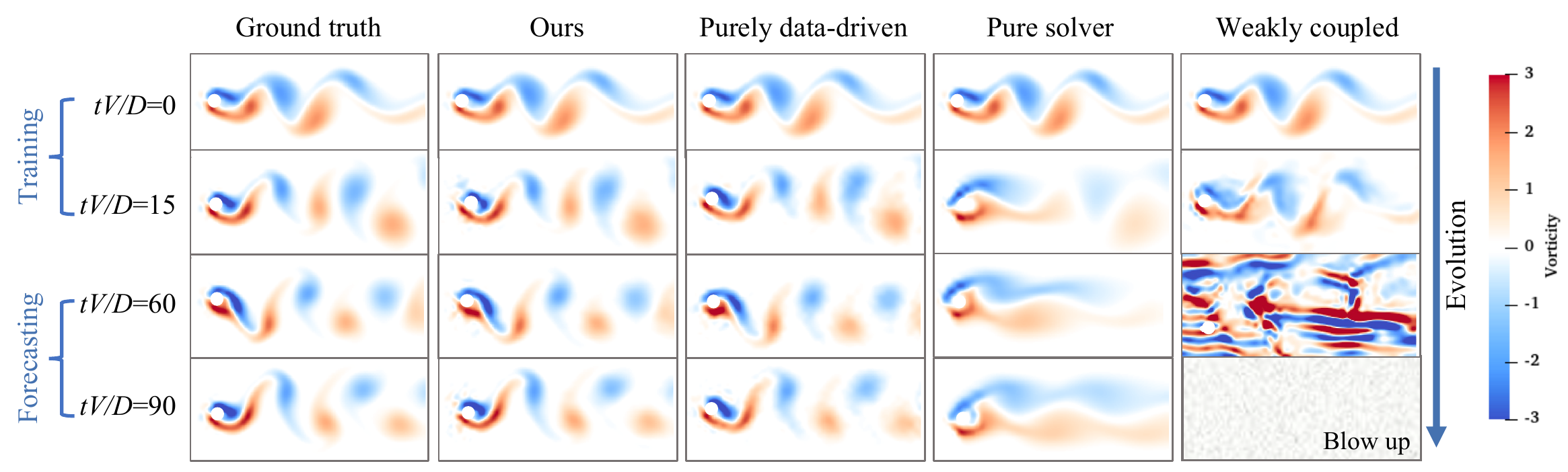}
\caption{Comparison of spatio-temporal vorticity predictions by different neural models when spring stiffness and reduced velocity is set as $k=0.16, U_{r}=6.25$.} 
\label{fig:Training_vorticity}
\end{figure}

\begin{figure}[!t]
\centering
\includegraphics[width=0.90\textwidth]{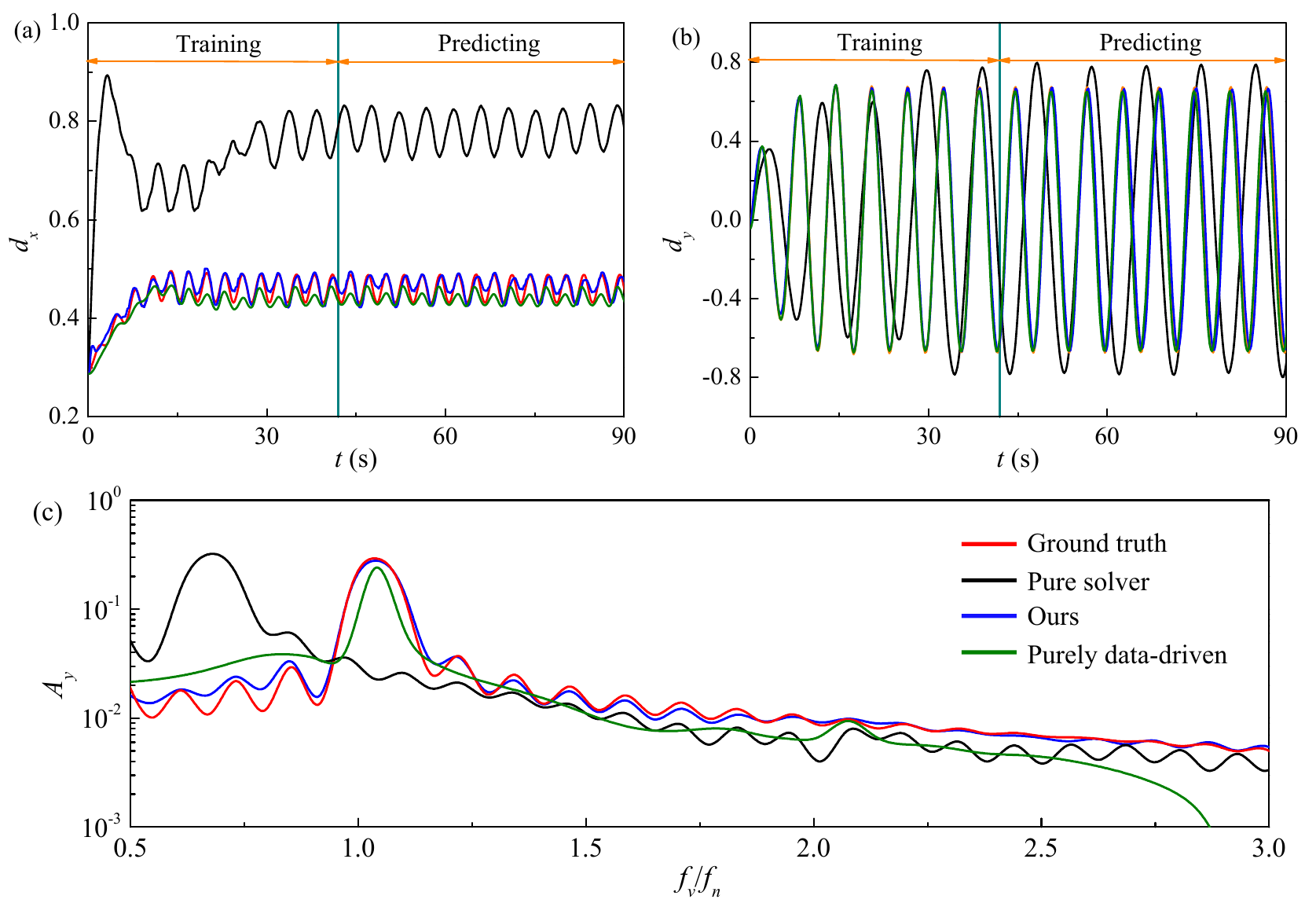}
\caption{Comparison of structural responses predicted by different models ($k=0.16$, $U_{r}=6.25$) : (a) displacement in streamwise direction; (b) displacement in transverse direction; (c) the spectra of transverse displacement with respect to non-dimensional vortex shedding frequency $f_v$. The weakly coupled model results are not presented as the simulation diverges} 
\label{fig:Training_structure}
\end{figure}

To evaluate the time-forecasting capabilities of the neural models, they are trained within the time window $tV/D = [0, 40]$. The performance of these trained models is then assessed by providing the initial conditions and generating predictions by a model rollout from $tV/D = 0$ to $tV/D = 90$. Specifically, the model rollout within the window $tV/D = [0, 40]$ represents inference in the training region, while the rollout within the window $tV/D = [40, 90]$ corresponds to time-forecasting. A comparison of the results between our proposed hybrid neural solver with the baseline models is presented in Fig.~\ref{fig:Training_vorticity} and Fig.~\ref{fig:Training_structure}. The results demonstrate that the proposed hybrid neural solver accurately forecasts the structural responses and vortex modes in a long time span using limited amount of training data. The vortex modes transition from a developing state to a steady `2S' mode, which is precisely captured by the our hybrid neural model ($2^\mathrm{rd}$ column of Fig.~\ref{fig:Training_vorticity}). These results are consistent with the ground-truth ($1^\mathrm{st}$ column of Fig.~\ref{fig:Training_vorticity}) and published findings \cite{wan2016suppression}. 
In contrast, a pure numerical FSI solver with the same spatio-temporal resolution struggles to capture the true vortex dynamics, as demonstrated in the $4^{\text{th}}$ column of Fig.~\ref{fig:Training_vorticity}. In this case, most flow features are damped out by the large numerical dissipation, which is a consequence of the very low spatio-temporal resolution used in learning-based neural solvers. Similarly, the predictions of structure responses (i.e., streamwise and transverse displacement $d_x$, $d_y$) by the pure solver (green) also show significant deviations from the ground truth (red), as illustrated in Fig.~\ref{fig:Training_structure}, where the predictions of the proposed hybrid model (blue) almost overlap with the reference. 

Note that the models here are tested on the same parameters ($k=0.16$, $U_{r}=6.25$) as seen during training, with the aim of exploring their performance in temporal forecasting. For testing on "seen" parameters, the temporal forecasting performance of the purely data-driven model is also reasonably good. The good performance of the purely data-driven model also benefits from the repetitive nature of VIV flows. Both the accuracy of its predicted flow fields and structural responses are much better than those of the pure numerical solver. However, compared to the proposed differentiable hybrid neural model, the purely data-driven model is less accurate. First, the vortex patterns in the far-wake region predicted by the purely data-driven model are not smooth and contain notable noises, as shown the $3^{\text{rd}}$ column of Fig.~\ref{fig:Training_vorticity}. Moreover, the streamwise displacements are underestimated and high-frequency features of transverse displacements are missed, as shown in Fig.~\ref{fig:Training_structure}. In particular, from the spectral analysis of displacement predictions (Fig.~\ref{fig:Training_structure}a), a notable discrepancy can be seen between the purely data-driven results and the ground true, while our hybrid neural model prediction is almost identical to the ground truth.    



The proposed hybrid neural model outperforms both the purely data-driven model and the pure numerical FSI solver in time-forecasting capabilities, thanks to the integration of discretized physics of FSI dynamics within a differentiable programming framework. Although the weakly coupled model also incorporates discretized governing PDEs and significantly reduces training loss, it experiences instability and rapid divergence during testing, as seen in the last column of Fig.~\ref{fig:Training_vorticity}. This highlights the severe instability issue of NN-embedded solvers, trained offline using \textit{a priori} loss criteria, emphasizing the advantages of the proposed differentiable hybrid neural model, which is trained end-to-end to meet \textit{a posteriori} criteria with gradients back-propagated over the entire solution trajectory.

To further examine the error propagation in time for all models, the relative errors $\epsilon^t$ of model prediction at each time step $t$ is computed as follows, 
\begin{equation}
  \epsilon^t=\frac{1}{N_\lambda} \sum_{i=0}^{N_{\lambda}}  \left(\frac{\lVert f(\mathbf{\widetilde{u}}^t, \lambda_i; \boldsymbol{\theta}^*)-\mathbf{u}^t_d \rVert _{2}}{\lVert \mathbf{u}^t_d(\lambda_i) \rVert _{2}}+\frac{\lVert \mathbf{w}^t-\mathbf{w}^t_d \rVert _{2}}{\lVert \mathbf{w}^t_d(\lambda_i)\rVert _{2}} \right) 
  \label{eq:relative error}
\end{equation}
where $N_\lambda$ is the size of parameter ensemble; $f$ represents the trained neural FSI solver with the optimized parameters $\boldsymbol{\theta}^*$; $\mathbf{w}$ is the structural responses calculated by the neural model; $\mathbf{u}^{t}_d$ and $\mathbf{w}^{t}_d$ are the ground-truth data.
\begin{figure}[!ht]
\centering
\includegraphics[width=0.85\textwidth]{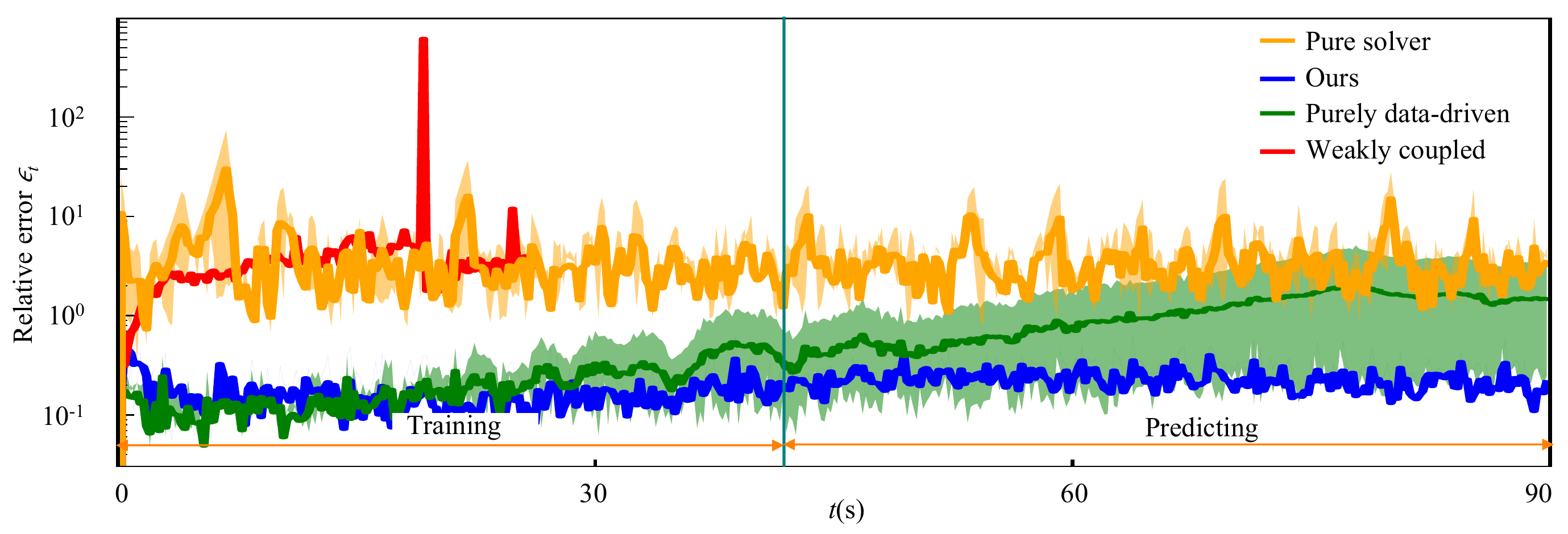}
\caption{The time evolution of the relative prediction errors from different models, where the thick lines indicated the mean value and the shaded regions represents the error scattering.}
\label{fig:Training_error}
\end{figure}
\begin{table}[!ht]
    \centering    
    \caption{The average relative errors of predictions for different models}
    \footnotesize
    \begin{tabular}{c c c c c}
    \toprule[1.5pt]
        $\bar{\epsilon}$ & Diff-hybrid neural (Ours) & Purely data-driven & Pure solver & Weakly coupled \\  \midrule
        Fluid & $\mathbf{0.16}$ & $0.67$ & $3.76$ & $\infty$  \\
        Structure & $\mathbf{0.02}$ & $0.11$ & $2.49$ & $\infty$ \\
    \bottomrule[1.5pt] 
    \end{tabular}
    \label{tab:training_error}
\end{table}
As illustrated in Fig.~\ref{fig:Training_error}, the relative error of the pure solver (yellow) remains orders of magnitude higher from the very beginning due to the low spatio-temporal resolution. The rollout error of the weakly coupled neural solver (red) rapidly grows even in the training range and quickly blows up at $t = 28$s. For the purely data-driven solver (green), the rollout error is initially low within the training range ($0 < t < 40$s) but quickly grows by an order of magnitude in the forecasting range ($t > 40$s) due to error accumulation, which is well-known for most data-driven time-forecasting models. In contrast, our proposed differentiable hybrid neural model (blue) maintains an impressively low error, even when extrapolating twice the length of the training range. Moreover, the error scattering is significantly reduced compared to the purely data-driven baseline, indicating the great robustness of the proposed model. Finally, the averaged relative errors over the entire rollout trajectory are presented in Tab.~\ref{tab:training_error}, demonstrating the notable superiority of the proposed model over other baselines.

\subsection{Flow-induced deformation (FID) of flexible structure}
To thoroughly showcase the exceptional performance of the proposed hybrid neural FSI solver across various FSI scenarios, we model the FID of a flexible plate as depicted in Fig.~\ref{fig:overview scheme}. In contrast to the first case, this case exhibits less periodic flow patterns and structural responses due to the added structural flexibility. We employ a ConvLSTM with identical architecture and hyperparameters as used in the first case to build this hybrid neural model. As discussed in Sec.\ref{sec:case}, the vertical flexible plate primarily experiences streamwise deformations with the first three vibration modes. For this demonstration, we select the most complex periods, during which the plate vibrates in the third mode, to highlight the advantages of our approach. The model undergoes training within the time window $tV/L = [0, 10]$, and its performance is evaluated afterwards by supplying the initial conditions and generating predictions through the model rollout spanning from $tV/D = 0$ to $tV/D = 30$. Namely, the model rollout within the window $tV/D = [0, 10]$ represents inference in the training range, while the rollout within the window $tV/D = [10, 30]$ corresponds to time-forecasting.

\begin{figure}[]
\centering
\includegraphics[width=1.0\textwidth]{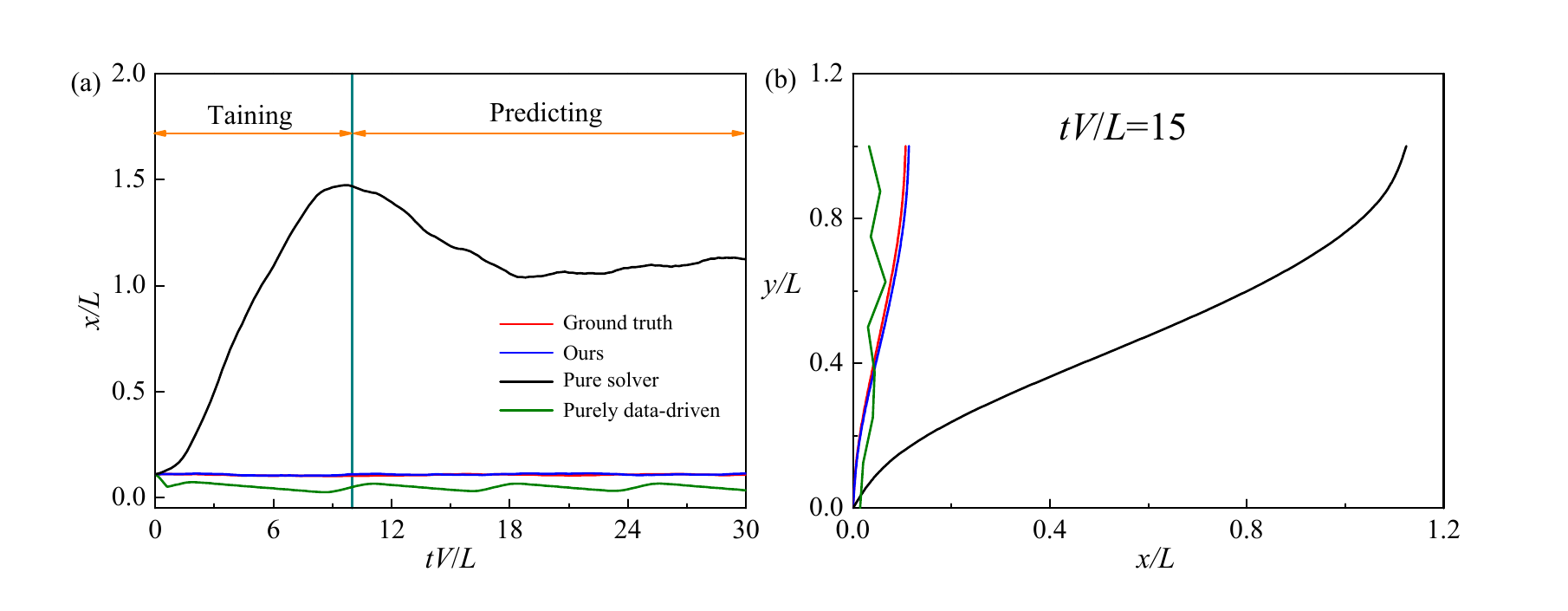}
\caption{Structural responses of different models in time sequence: (a) the temporal streamwise displacements of the free-end of the flexible plate; (b) the vibration modes of the plate at $tV/L=15$.}
\label{fig:flexible_structure}
\end{figure}
\begin{figure}[!ht]
\centering
\includegraphics[width=1.0\textwidth]{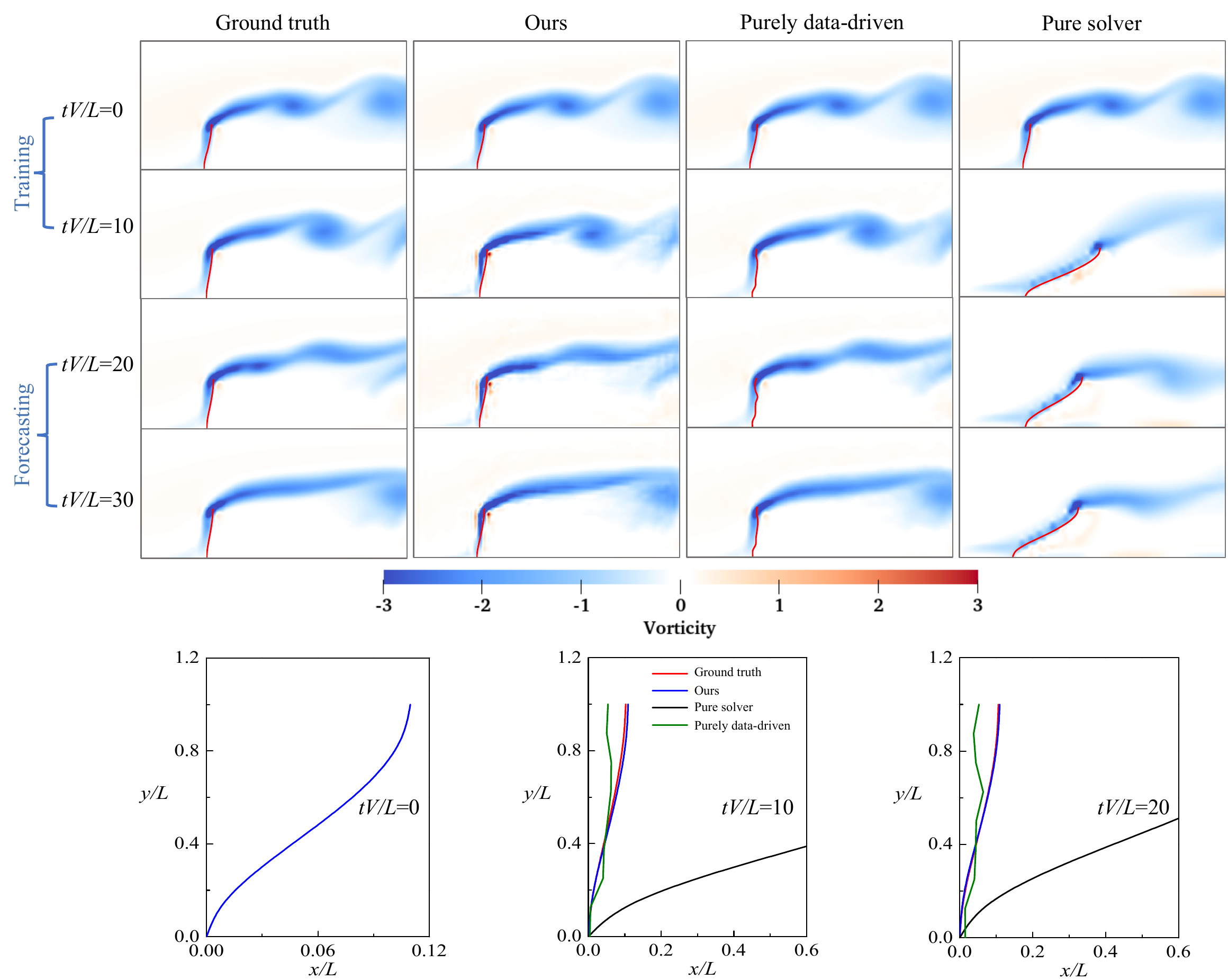}
\caption{Comparison of the vorticity and displacements of different models for flexible structure, where the flexible structure vibrating in the third mode is represented by red line.}
\label{fig:flexible_vorticity}
\end{figure}
The results are presented in Fig.~\ref{fig:flexible_structure} and Fig.~\ref{fig:flexible_vorticity}, excluding the analysis for the weakly coupled model, as it again quickly diverges during testing. Taking the displacements at the free end of the flexible plate as an example, it is evident that the proposed differentiable hybrid neural model accurately predicts time history of free-tip displacements, which agree well with the ground truth (see Fig.~\ref{fig:flexible_structure}(a)). Similar to the first case, the pure coarse FSI solver struggle to capture the structural responses, as it significantly overestimates the free-end displacement. Although the purely data-driven model appears to provide reasonable displacement predictions in magnitude, the vibration modes shown in Fig.~\ref{fig:flexible_structure}(b) reveal that the results are completely nonphysical, suggesting that the black-box model without any physics components performs worse in more complex FSI scenarios with flexible bodies. In contrast, the proposed hybrid neural FSI model solves the structural response based on physics-based governing equations, ensuring that the results are not only accurate but also physically consistent. In this case, both the hybrid neural model and purely data-driven model are capable of capturing the non-periodic flow patterns, as shown in Fig.~\ref{fig:flexible_vorticity}.


\section{Discussion}
\label{sec:discussion}

\subsection{Generalizability over unseen material parameters}
\label{sec: generalizability}
This subsection aims to assess the generalizability of the proposed model by evaluating its prediction performance across various testing points in the parameter space. In particular, we investigate the model's ability to make accurate predictions for unseen material parameters in VIV, such as spring stiffness, that were not included in the training dataset. To this end, we feed the trained neural FSI model with initial conditions for different values of spring stiffness, i.e., $k=0.18$ and $k=0.14$, and then roll out the model to predict the system behavior for a long time horizon.

\begin{figure}[]
\centering
\includegraphics[width=1.0\textwidth]{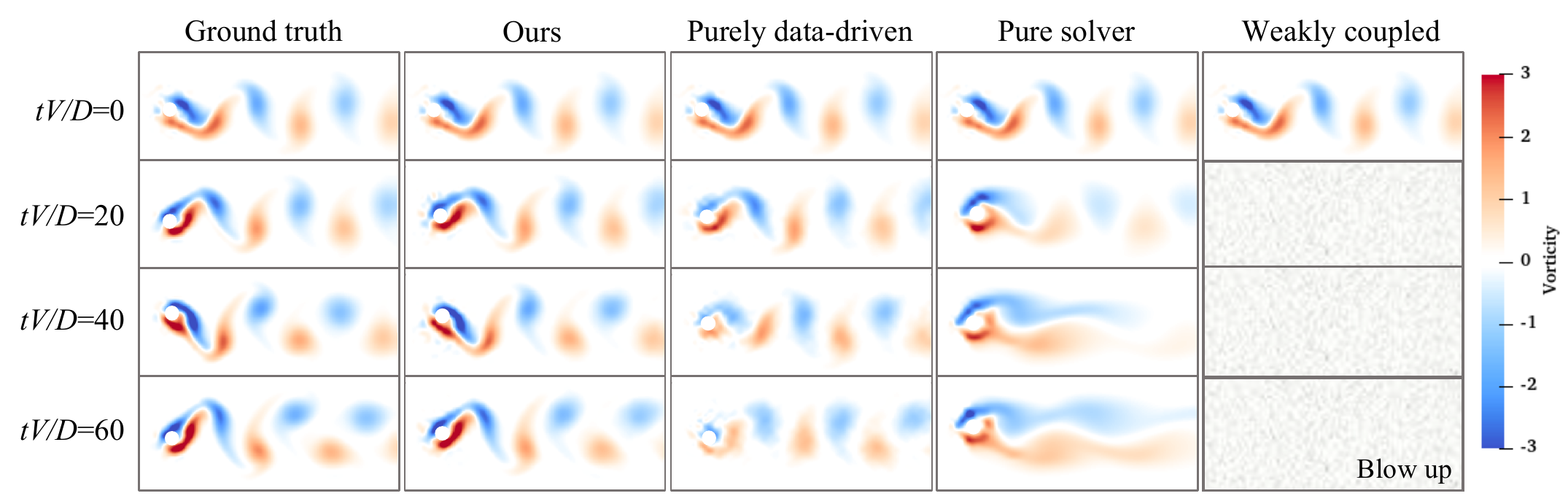}
\caption{Comparison the vorticity of different models in time sequence for unseen testing parameter $k=0.18$($U_{r}=5.56$).}
\label{fig:Testing_vorticity_0.18}
\end{figure}

\begin{figure}[]
\centering
\includegraphics[width=0.95\textwidth]{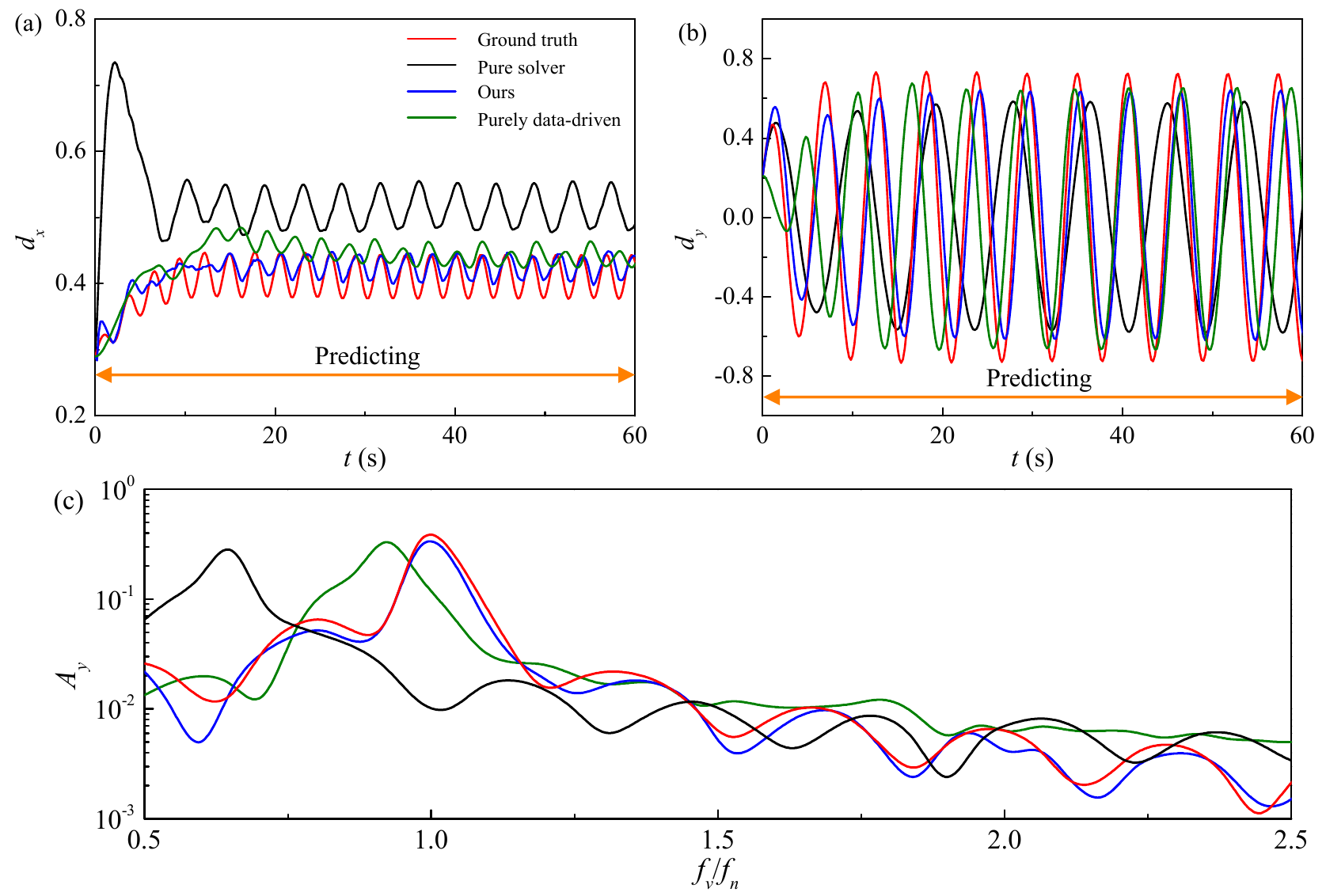}
\caption{The time-series structural responses for unseen parameter $k=0.18$($U_{r}=5.56$): (a) the displacement in streamwise direction; (b) the displacement in transverse direction; (c) the spectra of transverse displacement. The results of weakly physics-informed model are not included here due to the entire solver directly blows up for unseen parameters.}
\label{fig:Testing_structure_k0.18}
\end{figure}

We consider an extrapolated stiffness parameter $k=0.18$ as an example. The predicted flow fields at different time snapshots are presented in Fig.~\ref{fig:Testing_vorticity_0.18}, and the corresponding temporal displacement history of the cylinder from $tV/D = 0$ to $tV/D = 60$ is plotted in Fig.~\ref{fig:Testing_structure_k0.18}. The neural FSI solver successfully captures the development of vortex modes from $tV/D=0-20$ to $tV/D=40-60$, aligning well with the ground truth. This accuracy extends to the simulation of structural responses, covering both the transition and stable cyclic phases. These results indicate that the trained hybrid neural model exhibits excellent predictive capability, even when material parameters fall outside the training set. However, the pure solver, as expected, completely fails to capture the accurate FSI dynamics.
When faced with scenarios involving new material parameters outside the training set, the other data-driven baseline models encounter greater difficulties. As shown in Fig.~\ref{fig:Testing_vorticity_0.18}, the weakly coupled neural model rapidly diverges even in the first a few rollout steps. The purely data-driven model, though had reasonable predictions within the training time range as shown in the previous case, has a hard time to accurately capture both the fluid dynamics and structural responses from the first a few rollout steps, given new stiffness parameters. The predicted vortex modes is inaccurate in near wake region and largely damped out in the far wake region (see the $3^{\rm{rd}}$ column of Fig.\ref{fig:Testing_vorticity_0.18}). The inaccurate vortex shedding frequency also lead to large discrepancy between the structural response predictions by the purely data-driven model and the ground truth, as depicted in Fig.~\ref{fig:Testing_vorticity_0.18}.  

\begin{figure}[!ht]
\centering
\includegraphics[width=0.85\textwidth]{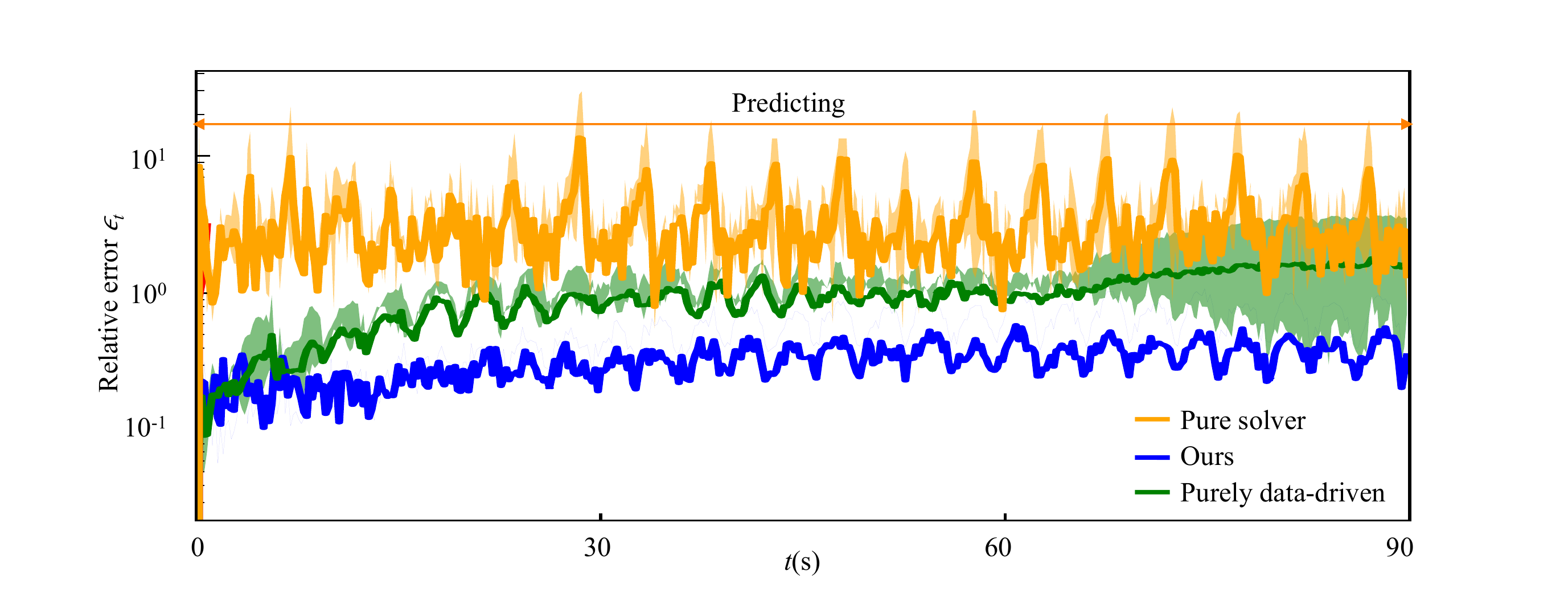}
\caption{The time evolution of the relative prediction errors from different models, where the thick lines indicated the mean value and the shaded regions represents the error scattering across all testing parameters.}
\label{fig:Testing_error}
\end{figure}   
\begin{table}[!ht]
    \centering    
    \caption{The averaged relative errors of predictions for different models (unseen stiffness)}
    \footnotesize
    \begin{tabular}{c c c c c}
    \toprule[1.5pt]
        $\bar{\epsilon}$ & Diff-hybrid neural (Ours) & Purely data-driven & Pure solver & Weakly coupled \\  \midrule
        Fluid & $\mathbf{0.21}$ & $1.01$ & $4.00$ & $\infty$  \\
        Structure & $\mathbf{0.05}$ & $0.64$ & $2.44$  & $\infty$\\
    \bottomrule[1.5pt] 
    \end{tabular}
    \label{tab:testing_error}
\end{table}
The time evolutions of relative prediction errors from various models are compared in Fig.\ref{fig:Testing_error}, and the average error values over the entire rollout trajectory are summarized in Table\ref{tab:testing_error}. It is evident that the prediction errors of our proposed model are significantly lower than those of other baseline models or the pure numerical solver with the same spatio-temporal resolution. Notably, there is virtually no error accumulation in the proposed hybrid neural model, in stark contrast to the purely data-driven model, whose prediction error continuously grows since in beginning of the model rollout. 

\begin{figure}[!ht]
\centering
\includegraphics[width=0.85\textwidth]{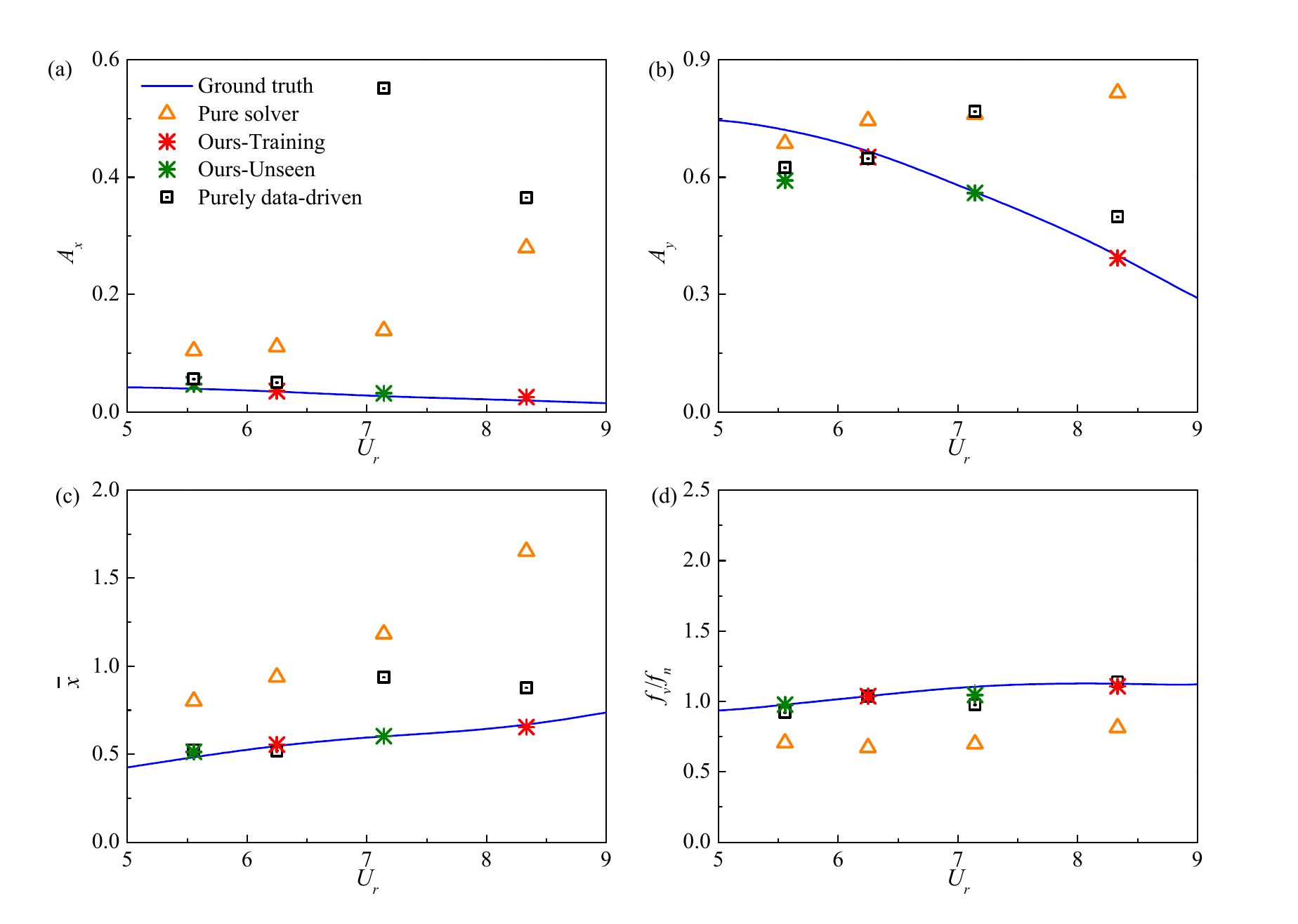}
\caption{Comparison of the prediction results in the parameter space: (a) the root-mean-square (RMS) of fluctuated displacement in streamwise direction; (b) the RMS of displacement in transverse direction; (c) the mean displacement in streamwise direction; and (d) the normalized vortex-shedding frequency.}
\label{fig:all parameter space}
\end{figure}

The statistical quantities of model predictions for two training parameters and two unseen parameters ($k$ or $U_r$) are summarized in Fig.\ref{fig:all parameter space}. The proposed neural solver accurately predicts the streamwise displacements, including both fluctuating and mean components for both interpolated ($k=0.14$, $U_r=7.14$) and extrapolated ($k=0.18$, $U_r=5.56$) parameters, as shown in Figs.~\ref{fig:all parameter space}(a) and~\ref{fig:all parameter space}(c). The transverse vibration amplitudes are perfectly predicted for the parameter interpolated between the training parameters; however, they are slightly underestimated for the extrapolated parameters. 

Based on the aforementioned analysis, the proposed hybrid neural solver demonstrates the ability to accurately predict FSI dynamics for unseen parameters, exhibiting good robustness and generalizability. It surpasses existing data-driven~\cite{gao2022quasi} and hybrid models~\cite{gupta2022hybrid, bukka2021assessment} for FSI, which are limited to forecasting responses in time sequences for trained parameters only. Furthermore, this analysis reveals that the sequential learning strategy enabled by differentiable programming surpasses the next-step net in learning complex spatio-temporal FSI dynamics. In addition to employing the sequential net (ConvLSTM), the discretized physics in the current neural model are formulated in a recurrent form, which can be interpreted as a sequence-net with non-trainable parameters predefined by classic numerical methods. Consequently, the proposed neural FSI model is capable of predicting the temporal coherence of two-way fluid-structure interactions in long-term rollouts.

\subsection{Offline training cost and online inference speedup}
\begin{figure}[!ht]
\centering
\includegraphics[width=10cm]{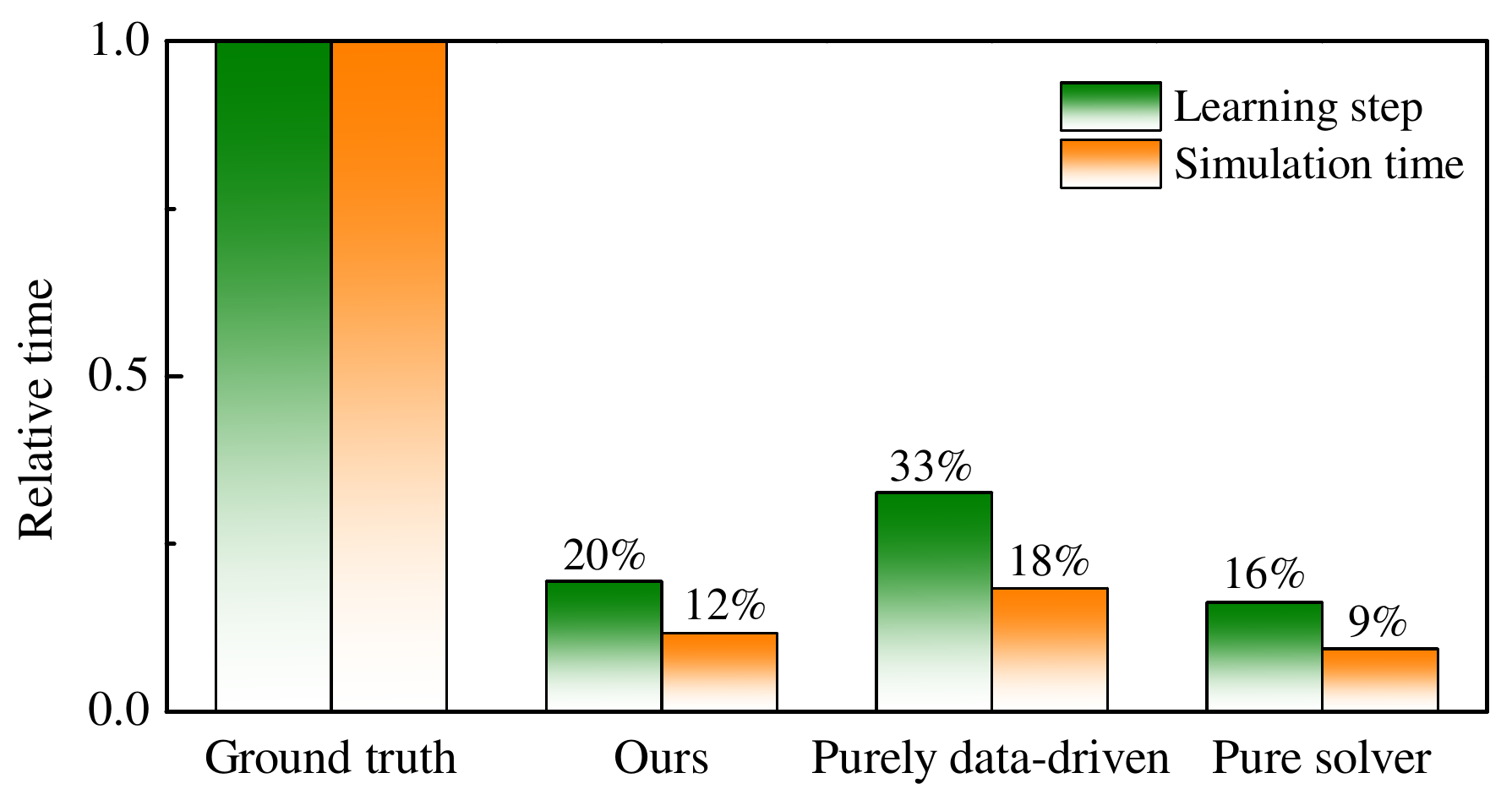}
\caption{Relative wall time among different models to evaluate the acceleration of neural FSI solver. The ground truth data is computed on $512 \times 256$ grids. The Neural models are operated on $64 \times 32$ grids, which is 8$\times$ coarsen in each dimension. The learning step of the neural solver is twice as numerical time step fine-grid, high-fidelity FSI simulation.}
\label{fig:cost-time}
\end{figure}

As shown in Fig.~\ref{fig:cost-time}, we assess the running time of various models during the online prediction phase, also known as inference. The inference cost of our trained hybrid neural FSI model for a single learning step is a mere $20\%$ of that of a numerical FSI solver on fine grids for one numerical time step. Considering that each learning step typically covers multiple numerical time steps, the proposed neural model offers notable speedup for long time-span predictions. For example, as demonstrated by the orange bars in Fig.~\ref{fig:cost-time} showing the relative cost for 2D VIV cases, the hybrid neural model can accelerate computation by 8.3 times when simulating the same physical time length. This acceleration is attributed to the use of spatially coarser grids and a larger time stepping interval, which is twice the numerical time step. 
In certain situations, a strong FSI coupling scheme and small numerical steps are required due to the numerical instability caused by the structure's flexibility. The strong coupling scheme also necessitates sub-iterations to solve structural responses in each time step. This implies that the computation cost of high-resolution simulations will be further increased considerably, especially for multi-degree-of-freedom structures. However, once trained on limited amount of high-resolution data, the optimized hybrid neural model can maintain a simple weak coupling scheme and rollout with a larger learning step. As a result, the speedup provided by the hyrid neural model can become even more significant.




The cost of training neural models (using an RTX A6000 GPU) is comparable to conducting a single high-fidelity FSI simulation (on 8 CPUs) in terms of wall time. However, the training represents a one-time offline expense, and the marginal cost for online evaluation of the trained model is very low. This makes it an ideal solution for scenarios that require long-term forecasting and numerous model queries within the parameter space.

For example, nonlinear FSI cases often necessitate simulations across extended trajectories to obtain reliable statistics. In such cases, the trained neural model can be employed to forecast long-term dynamics, provided that accurate initial conditions can be determined through short-term numerical simulations. In this context, the costs associated with training and inference will be significantly lower than those of performing high-fidelity simulations directly. Moreover, hybrid neural modeling can greatly enhance design optimization and uncertainty quantification within the parameter space, as the trained model can be generalized to new parameters without necessitating re-training.

We must acknowledge that the current training cost of the neural FSI solver is higher than that of a purely data-driven model. For Seq2Seq training with 300 rollout steps, the training cost of the neural solver is 7.7 times that of the purely data-driven model (as measured by wall time). This increased cost is due to the memory-intensive and time-consuming back-propagation process, which traces gradients through all the differentiable modules. However, this training overhead can be offset by the significantly enhanced inference performance and generalizability, as demonstrated in Sec.~\ref{sec: generalizability}.   


\section{Conclusion}
\label{sec:conclusion}
In this paper, we presented a differentiable hybrid neural model for efficiently simulating FSI problems. By integrating the numerically discretized FSI physics using FVM and IBM with deep sequential neural networks using differentiable programming, we enabled end-to-end, sequence-to-sequence training of the entire hybrid model for long-term model rollouts. 

The merits and effectiveness of the neural FSI solver have been demonstrated through two typical FSI benchmarks, involving the dynamics of a rigid and flexible body, respectively. The flow patterns, fluid forces, and structural responses can be accurately modeled over long time steps with limited training datasets. In comparison to black-box neural network models, the proposed hybrid neural solver exhibits significant superiority in terms of generalizability, robustness, and prediction speed. It also overcomes numerical instability issues encountered by the weakly coupled hybrid model, as the fully differentiable architecture enables Seq2Seq training based on the \textit{a posteriori} criteria. The discretized governing PDEs are integrated as a part of a LSTM-based sequence net, which significantly mitigated the error accumulation issue, that is very common in most data-driven forecasting models.

Although the offline training process is somewhat time-consuming, which is just one time effort. Once the model is trained, the online inference speed of is considerably faster than that of classical numerical methods to achieve the similar accuracy. The speedup can be further improved by 8.3 times for long-term forecasting. This suggests that the proposed neural solver is well-suited for many FSI applications requiring long-span simulation or repeated model queries, such as design optimization and uncertainty quantification. 

In conclusion, the seamless integration of deep learning and classical numerical techniques through the differentiable programming framework has successfully leveraged the advantages of both methods, resulting in a powerful approach for modeling complex FSI dynamics. This innovative method shows great promise in the development of next-generation data-enabled neural solvers that achieve a balance between accuracy and computational efficiency, making it an ideal tool for addressing the challenges posed by large-scale and long-duration FSI simulations.

 \section*{Acknowledgment}
 The authors would like to acknowledge the funds from Office of Naval Research under award numbers N00014-23-1-2071 and National Science Foundation under award numbers OAC-2047127. XF would also like to acknowledge the fellowship provided by the Environmental Change Initiative and Center for Sustainable Energy at University of Notre Dame. 

\section*{Compliance with Ethical Standards}
Conflict of Interest: The authors declare that they have no conflict of interest.

\appendix
\section{Governing equations for two-type solids}
\label{sec:append-equation}
The specific form of the governing equations for structure dynamics (Eq.~\ref{eq: vibration equation}) varies between rigid and flexible bodies. For rigid solids, where deformations are neglected, the equations can be simplified to mass-damper-spring systems, which correspond to vortex-induced vibration (VIV) phenomena. In this case, the matrices $\mathbf{M}$, $\mathbf{K}$, and $\mathbf{C}$ in Eq.~\ref{eq: vibration equation} are reduced to scalar values. The two-degree-of-freedom equation at time step $t$ can be rewritten as:
\begin{equation}
    m
    \left[
    \begin{array}{c}
         \ddot{w}_x  \\
         \ddot{w}_y 
    \end{array}
    \right]
    +
    k
        \left[
    \begin{array}{c}
         w_x  \\
         w_y 
    \end{array}
    \right]
    +
    c
        \left[
    \begin{array}{c}
         \dot{w}_x  \\
         \dot{w}_y 
    \end{array}
    \right]
    =
            \left[
    \begin{array}{c}
         F_D  \\
         F_L 
    \end{array}
    \right],
\label{eq: rigid-body vibration}
\end{equation}
where $m$ is the mass; $k=2m \cdot (2\pi f_{n})^{2}$, and $f_{n}$ is natural frequency; $c=2m\cdot2\pi f_{n} \cdot \xi_{s}$, where $\xi_{s}$ is structural damping ratio; $F^D=\sum_{i=1}^{n} q^D_{i}$ and $F^t_L=\sum_{i=1}^{n}q^D_{i}$ are drag and lift force, and $n$ is the total number of Lagrangian nodes.
For general flexible solids, the structure is modeled by assembling beam elements. Considering only the bending of beam elements in a two-dimensional domain, the consistent mass matrix $\mathbf{M}$ and stiffness matrix $\mathbf{K}$ for flexible solids are constructed. Since each single beam element consists of the $i^{th}$ and $(i+1)^{th}$ nodes, we can derive the local stiffness and mass matrix for an individual element:
 \begin{equation}
     \mathbf{K}^e=\frac{EI}{L_e^3}
     \left[
     \begin{array}{c c c c}
     12 & 6 L_e & -12 & 6L_e \\
     6L_e & 4L_e^2 & -6 L_e & 2L_e^2 \\
     -12 & -6 L_e & 12 & -6 L_e \\
     6L_e & 2L_e^2 & -6 L_e & 4 L_e^2
     \end{array}
     \right],
 \end{equation} 

  \begin{equation}
     \mathbf{M}^e=\frac{\mu_s L_e}{420}
     \left[
     \begin{array}{c c c c}
     156 & 22 L_e & 54 & -13L_e \\
     22L_e & 4L_e^2 & 13 L_e & -3L_e^2 \\
     54 & 13 L_e & 156 & -22 L_e \\
     -13L_e & -3L_e^2 & -22 L_e & 4 L_e^2
     \end{array}
     \right],
 \end{equation} 
where $L_e=z_{i+1}-z_i$. Here, the damping matrix is constructed based on Rayleigh damping~\cite{liu1995formulation}:
\begin{equation}
\mathbf{C}=\alpha\mathbf{M}+\beta\mathbf{K},
\end{equation}
where the coefficients $\alpha =2\xi_s \dfrac{\omega_i \cdot \omega_j}{\omega_i + \omega_j}$, $\beta =\dfrac{2\xi_s}{\omega_i + \omega_j}$. $\omega_i$ is the $ith$ natural circular frequency solved by:
 \begin{equation}
     (\mathbf{K}-\mathbf{\omega}^2\mathbf{M})\mathbf{\Phi}=0,
 \end{equation}
where $\mathbf{\omega}=diag[\omega_1 \dotsm \omega_i \dotsm \omega_j \dotsm \omega_n]$ is the eigenvalue, and $\mathbf{\Phi}$ is the eigenvector.

\section{Implementation and training details of hybrid neural model}
\label{sec:append-a}
\setcounter{figure}{0}

\subsection{Partitioned long-trajectory training}
\label{sec:trajectory}
The sequence-to-sequence training process is memory-intensive, especially when combined with differentiable physics components. To circumvent memory constraints, the entire trajectory in the current training is divided into muliple subchains. The mean squared error (MSE) loss for the complete trajectory is calculated as follows,
\begin{equation}
    \mathcal{L}(\boldsymbol{\theta})=\frac{1}{N} \sum_{t=0}^{N} ( f(\Tilde{\mathbf{u}}_{t}; \boldsymbol{\theta}) - \mathbf{u}_{t}^d )^{2},
    \label{eq:loss-a}
\end{equation}
where $f(\Tilde{\mathbf{u}}_{t}; \boldsymbol{\theta})$ is the output of the hybrid neural FSI model; $\mathbf{u}_{t}^d$ is the label data; and $N$ is the number of time steps. If the entire trajectory is divided into $n$ equal-time-span sub-trajectories, and each sub-trajectory includes $N_{sub}$ time steps. The loss $\mathcal{L}$ in Eq.\ref{eq:loss-a} can be calculated by:
\begin{equation*}
\begin{aligned}
    \mathcal{L} (\boldsymbol{\theta}) &=\frac{1}{N} \sum_{i=0}^{n} \sum_{t=0}^{N_{sub}} ( f(\Tilde{\mathbf{u}}_{t,i}; \boldsymbol{\theta}) - \mathbf{u}_{t,i}^d )^{2} \\
    & = \frac{N_{sub}}{N} \sum_{i=0}^{n} {\frac{1}{N_{sub}}  \sum_{t=0}^{N_{sub}} ( f(\Tilde{\mathbf{u}}_{t,i}; \boldsymbol{\theta}) - \mathbf{u}_{t,i}^d )^{2}} \\
    & = \frac{N_{sub}}{N} \sum_{i=0}^{n} \frac{1}{N_{sub}}  \lVert f(\Tilde{\mathbf{u}}_{t,i}, \boldsymbol{\theta}) - \mathbf{u}_{t,i}^d \rVert_{2}^{2}  \\
    & = \frac{N_{sub}}{N} \sum_{i=0}^{n} \mathcal{L} _{i}
\end{aligned}
\end{equation*}
Accordingly, the total gradients can be obtained written as,
\begin{equation}
    \frac{\partial \mathcal{L}}{ \partial \boldsymbol{\theta}} = \frac{N_{sub}}{N} \sum _{i=0}^{n} \frac{\partial \mathcal{L}_i}{ \partial \boldsymbol{\theta}},
\end{equation}
where $N_{sub}$ is the number of time steps in each sub-trajectory; and $n$ is the number of sub-trajectory. It is important to note that the final outputs and hidden states must be transferred to the subsequent sub-trajectory during training. The training algorithm for the proposed hybrid neural FSI model is presented in Algorithm~\ref{training}.
\begin{algorithm}
\footnotesize
\SetKwInput{KwInit}{Initialize}
\caption{Training algorithm for the differentiable hybrid neural FSI model} 
\KwData{Load label fluid and/or structure data}
\KwInit{Initialize neural networks and discretized physics}
$epoch \gets 0$
\While{$epoch \leqslant N$}{

    \For{$t \in [1,N_{sub}]$} {
        $\widetilde{\mathbf{u}}^*_t \gets \mathcal{F}(\mathbf{u}_{t-1}, \mathbf{x}_{t-1}) $ \Comment{Solve intermediate velocity} 
    
        $\mathbf{p}_t \gets \mathscr{F}(\widetilde{\mathbf{u}}^*_{t}, \mathbf{x}_{t-1}, \rm{state}; \boldsymbol{\theta}_1) $ \Comment{Predict pressure}
    
        $\widetilde{\mathbf{u}}_t  \gets \widetilde{\mathbf{u}}^*_t-\frac{\nabla \mathbf{p}_t \Delta t}{\rho} $  \Comment{Get low-resolution velocity}
    
        $\mathbf{u}_t  \gets \mathscr{F}(\widetilde{\mathbf{u}}_t, \mathbf{x}_{t-1},  \rm{state}; \boldsymbol{\theta}_2)$  \Comment{Predict velocity} 
        
        $ \mathbf{x}_t \gets   \mathcal{G} (\mathbf{u}_t, \mathbf{x}_{t-1}, \lambda) $ \Comment{Directly solve structural responses}
        }
    $\mathcal{L} \gets \lVert \rm{predict} - \rm{label} \rVert_{L_2}^{2}$ \

    $\boldsymbol{\theta}_1 \gets \boldsymbol{\theta}_1-l \frac{\partial \mathcal{L}}{ \partial \boldsymbol{\theta}_1}$ \
    
    $\boldsymbol{\theta}_2 \gets \boldsymbol{\theta}_2-l \frac{\partial \mathcal{L}}{ \partial \boldsymbol{\theta}_2}$ \
       
}\label{training}
\end{algorithm}

\subsection{Downsampling of high resolution ground-truth data}
\label{sec: downsample}

\begin{figure}[H]
\centering
\includegraphics[width=8cm]{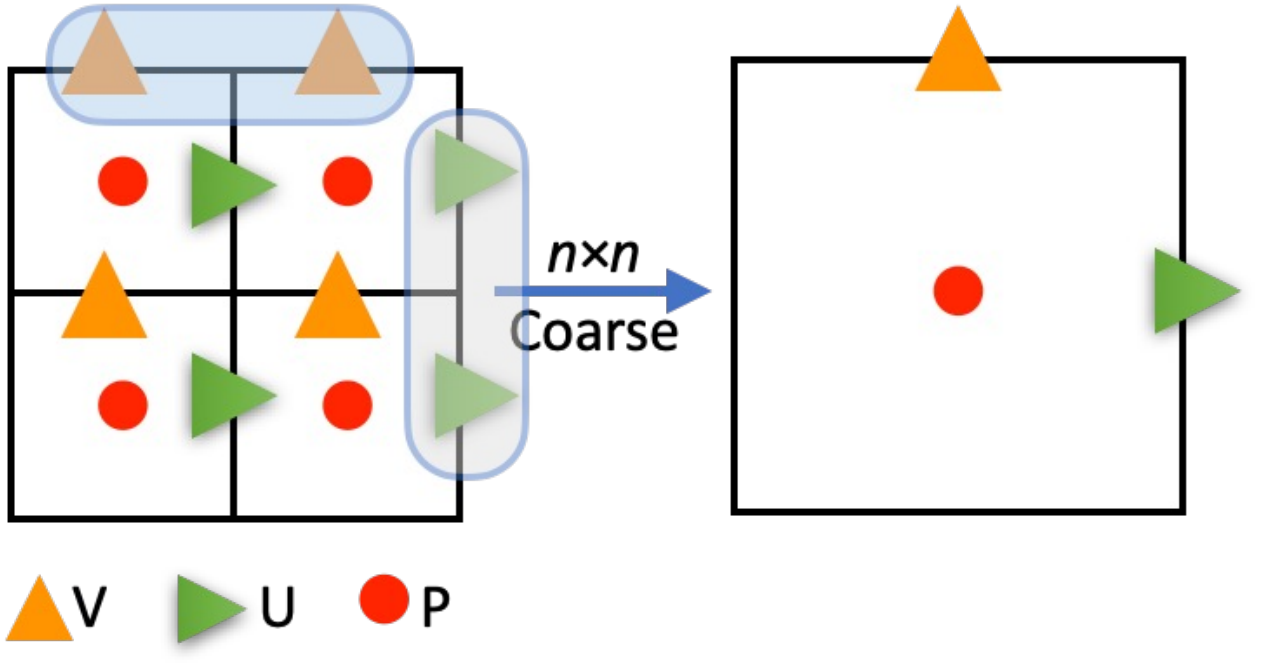}
\caption{Scheme of downsampling staggered variables to a 2x coarser grid. Velocity components on the fine-mesh located on faces of a coarse-mesh unit-cell (as indicated by shaded regions) are averaged to compute the velocities on the coarse-grid. The pressures located in the fine-mesh unit-cell centre are averaged to compute the pressure of a coarse mesh. Velocity components in the interior of the unit-cell are discarded. }
\label{sfig:down sample scheme}
\end{figure}

\subsection{Hyperparameters of the trainable ConvLSTM block}
\label{sec: Hyperparameters}
In the hybrid neural FSI model, all ConvLSTM blocks are identical consists of five layers, with channels configurations of $[32,64,64,64,32,1]$ and a trainable $3 \times 3$ convolutional kernel. The activation function is a rectified linear unit (ReLU), with the exception of the last layer, which does not have an activation function. The optimization settings are listed as below,
\begin{itemize}
    \item {Initial learning rate = $10^{-4}$}
    \item {Optimizer = Adam}
    \item {Scheduler = CosineDecaySchedule (alpha=$10^{-9}$)}
\end{itemize}
The weighting parameters of the fluid and solid loss terms $\alpha$ and $\beta$ are set as one.

\section{Baseline data-driven neural models}
\label{sec:append-b}
\setcounter{figure}{0}  
\begin{figure}[H]
\centering
\includegraphics[width=1.0\textwidth]{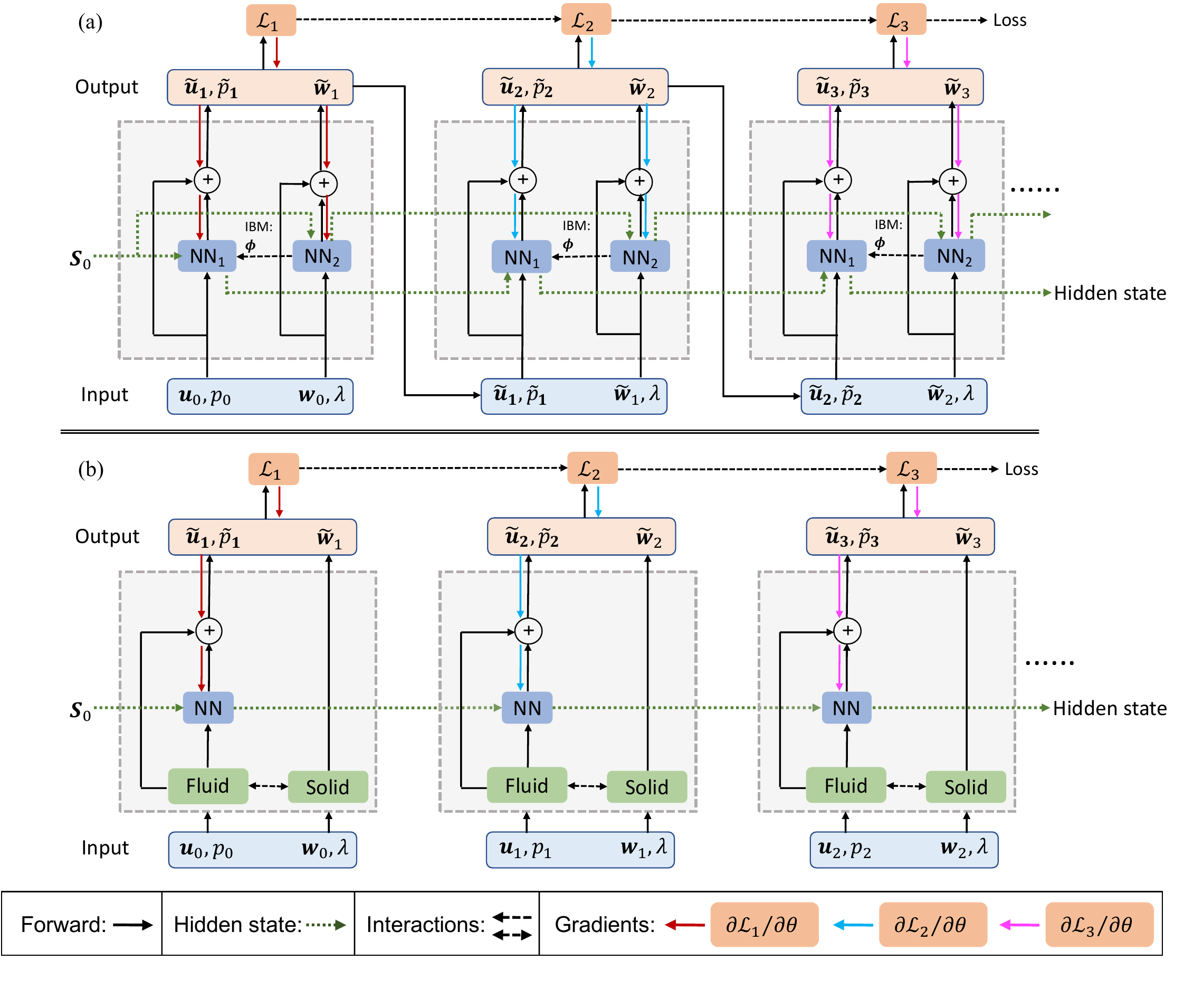}
\caption{(a) Purely data-driven model, where the known physics are completely removed and the dynamics of fluid and solid between two time steps are respectively mapped by LSTM neural networks. The fluid and solid nets are connected by IBM masks. 
(b) The weakly coupled model, where the discretized physics are preserved but the gradients cannot be back-propagated through them, equivalent to an offline network trained in teacher forcing manner. The trainable neural architecture remain the same as the proposed neural model.}
\label{sfig:base line}
\end{figure}

\begin{figure}[H]
\centering
\includegraphics[width=8cm]{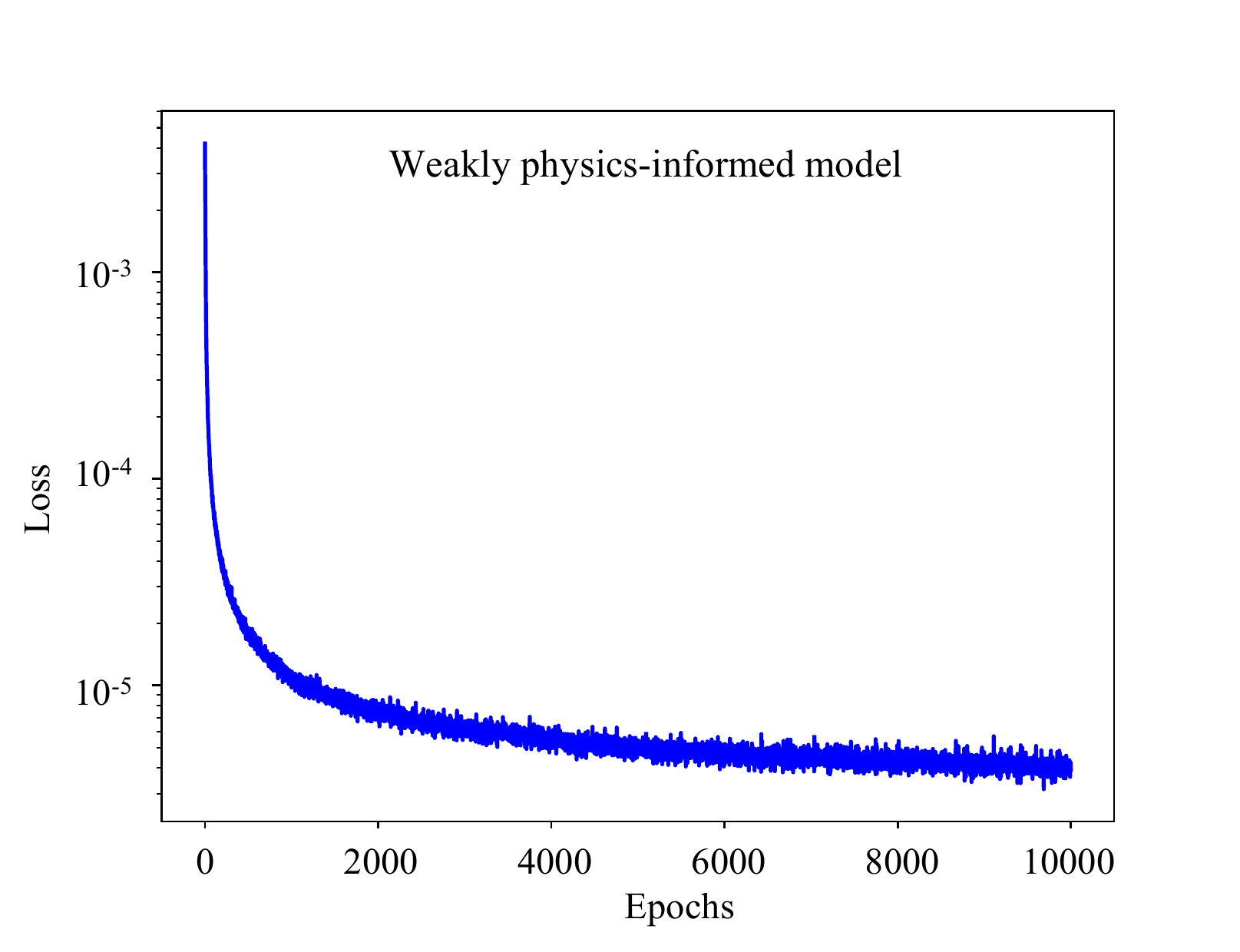}
\caption{The training loss for weakly coupled hybrid neural model.}
\label{sfig:loss of weakly couple}
\end{figure}

\begin{figure}[H]
\centering
\includegraphics[width=12cm]{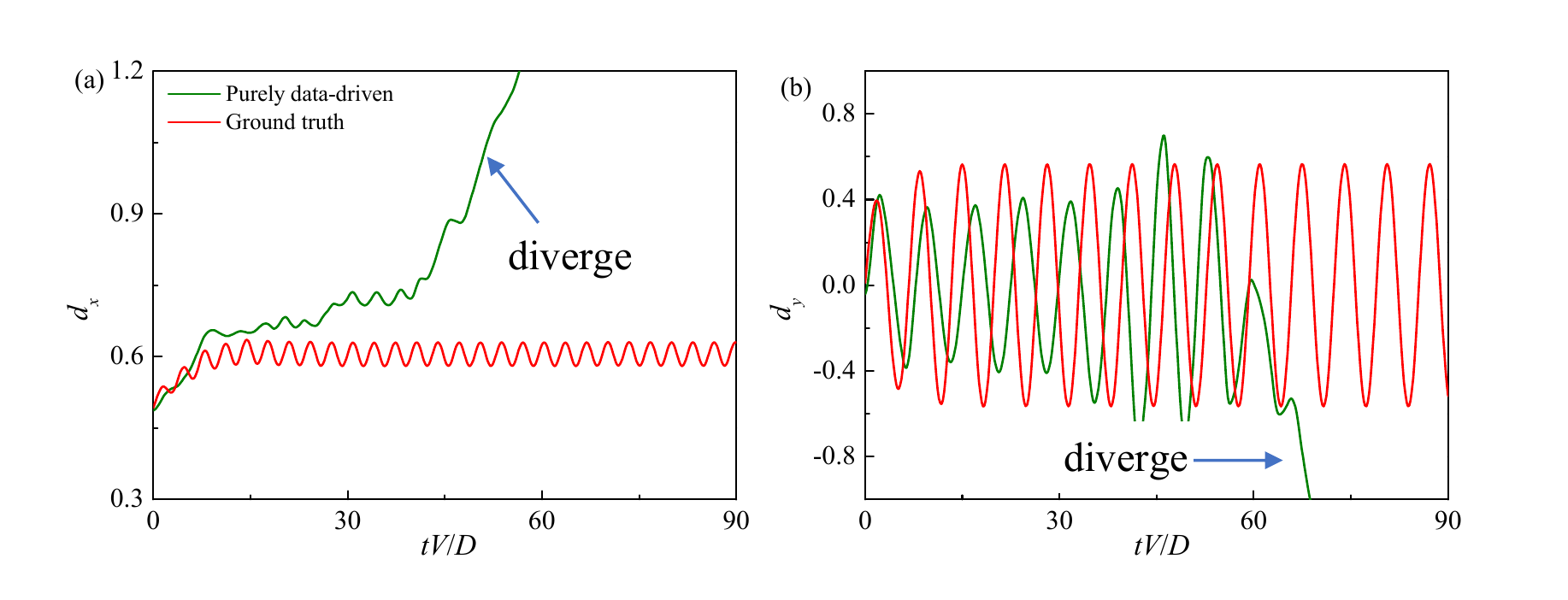}
\caption{The diverged structural responses predicted by purely data-driven model for $k=0.14$, $U_{r}=7.14$: (a) displacement in streamwise direction; (b) displacement in transverse direction.}
\label{sfig:diverge of black box}
\end{figure}



\bibliographystyle{elsarticle-num}
\bibliography{ref,own-ref}

\end{document}